\documentclass[a4paper,11pt]{article}
\pdfoutput=1
\usepackage{jheppub}
\usepackage[utf8]{inputenc}
\usepackage[english]{babel}
\usepackage{amsmath}
\usepackage{amsfonts}
\usepackage{amssymb}
\usepackage{anyfontsize}
\usepackage{listings}
\usepackage{lipsum}
\usepackage{datetime}
\usepackage{graphicx}
\usepackage{mathtools}
\usepackage{tensor}
\usepackage{mathrsfs}
\usepackage{dcolumn}
\usepackage{multirow}
\usepackage{braket}
\usepackage[T1]{fontenc} 
\usepackage[force]{feynmp-auto}

\usepackage{cleveref}

\crefformat{section}{\S#2#1#3} 
\crefformat{subsection}{\S#2#1#3}
\crefformat{subsubsection}{\S#2#1#3}

\newcommand{\bs}{\boldsymbol}

\newcommand{\mc}{\mathcal}

\usepackage{color}
\hypersetup{
    colorlinks=true, 
    pdfborder = {0 0 0.5 [3 3]},
    anchorcolor=black,
    citecolor=blue,
    linktoc=all,    
    linktocpage=true,
    linkcolor=blue,
    urlcolor=blue
}

\title{Post-Newtonian Waveforms from Spinning Scattering Amplitudes}

\author[a,b]{Yilber Fabian Bautista,}
\author[a,c,d]{Nils Siemonsen}

\affiliation[a]{Perimeter Institute for Theoretical Physics, Waterloo, ON N2L 2Y5, Canada}
\affiliation[b]{Department of Physics and Astronomy, York University, Toronto, Ontario, M3J 1P3, Canada.}
\affiliation[c]{Arthur B. McDonald Canadian Astroparticle Physics Research Institute, 64 Bader Lane, Queen's University, Kingston, ON K7L 3N6, Canada}
\affiliation[d]{Department of Physics and Astronomy, University of Waterloo, Waterloo, ON N2L 3G1, Canada}

\emailAdd{ybautistachivata@perimeterinstitute.ca}
\emailAdd{nsiemonsen@perimeterinstitute.ca}

\abstract{We derive the classical gravitational radiation from an aligned spin binary black hole on \textit{closed} orbits, using a dictionary built from the 5-point QFT scattering amplitude of two massive particles exchanging and emitting a graviton. We show explicitly the agreement of the transverse-traceless components of the radiative linear metric perturbations -- and the corresponding gravitational wave energy flux -- at future null infinity, derived from the scattering amplitude and those derived utilizing an effective worldline action in conjunction with multipolar post-Minkowskian matching. At the tree-level, this result holds at leading orders in the black holes' velocities and up to quadratic order in their spins. At sub-leading order in black holes' velocities, we demonstrate a matching of the radiation field for quasi-circular orbits in the no-spin limit. At the level of the radiation field, and to leading order in the velocities, there exists a one-to-one correspondence between the binary black  hole  mass and current quadrupole moments, and the scalar and linear-in-spin scattering amplitudes, respectively. Therefore, we show explicitly that waveforms, needed to detect gravitational waves from \textit{inspiraling} binary black holes, can be derived consistently, to the orders considered, from the classical limit of quantum \textit{scattering} amplitudes.}

\begin{document} 
\maketitle
\flushbottom
\newpage
\section{Introduction}

In the early days of general relativity, Einstein predicted the existence of gravitational waves \cite{1918SPAW154E} and cast the emission from a compact system into the, now famous, Quadrupole formula for gravitational radiation. A little while later, in a spectacular breakthrough the LIGO/Virgo Collaboration \cite{LIGOScientific:2016aoc} confirmed Einstein's prediction by directly detecting the gravitational waves emitted from a binary black hole (BBH). Higher order corrections to Einstein's Quadrupole formula in the context of the quasi-circular orbit general relativistic two-body problem -- needed to enable such detections -- have traditionally been obtained in the post-Newtonian (PN) \cite{Blanchet:2013haa,Futamase:2007zz} formalism, within numerical relativity \cite{Pretorius:2005gq} and black hole perturbation theory \cite{1973Teukolsky,Kokkotas:1999bd}, as well as models combining these approaches \cite{PhysRevD.59.084006,Buonanno:2000ef,Santamaria:2010yb}. More recently, however, efforts have been focused on the BBH scattering problem, in order to connect classical computations performed in the context of post-Minkowskian (PM) theory \cite{Damour_2016,Porto:2016pyg,Goldberger:2017vcg,Goldberger:2017ogt,Vines:2018gqi,Damour:2019lcq,Damour:2020tta,Kalin:2019rwq,Kalin:2019inp,Kalin:2020mvi,Kalin:2020fhe,Goldberger:2020fot,Brandhuber:2021kpo,Brandhuber:2021eyq}, with those approaches based on the classical limit of QFT scattering amplitudes \cite{Cheung:2020gyp,Cheung:2018wkq,Bern:2019nnu,Bern:2019crd,Bern:2021dqo,Bjerrum-Bohr:2018xdl,Cristofoli:2019neg,Bjerrum-Bohr:2019kec,Bjerrum-Bohr:2021vuf,DiVecchia:2020ymx,DiVecchia:2021ndb,Bern:2020buy,Chung:2019duq,Chung:2020rrz,Cachazo:2017jef,Guevara:2017csg,Guevara:2018wpp,Guevara:2019fsj,Aoude:2020onz,Bautista:2019evw, Blumlein:2020pyo, Blumlein:2020znm}. 

Until recently, the scattering amplitudes approach to the two-body scattering problem had mostly focused it's efforts in the conservative sector, although hints from soft theorems suggest that they can also be used to address the radiative sector \cite{Laddha:2018rle,Saha:2019tub,Ghosh:2021hsk}. 
The introduction of the Kosower, Maybee and O'Connell (KMOC) formalism \cite{Kosower:2018adc}, enabling the computation of classical observables directly from the scattering amplitude, proved to be extremely useful in determining radiative observables. Using this formalism, an amplitudes derivation of the waveform from hyperbolic, soft encounters was presented in \cite{Bautista:2019tdr}, encapsulating the gravitational memory content of the signal \cite{1987KPT}. In this same formalism, the computation of the full leading PM order radiated four-momentum was recently presented in \cite{Herrmann:2021lqe,Herrmann:2021tct}; these results were subsequently confirmed by other methods in \cite{DiVecchia:2021bdo,Bini:2021gat,Riva:2021vnj}. Simultaneously, using a worldline-QFT formalism \cite{Mogull:2020sak}, the computation of the  gravitational waveform valid for all values for the momentum of the  emitted graviton, was computed  in \cite{Jakobsen:2021smu} (see also \cite{,Mougiakakos:2021ckm}), and extended to include spin effects in \cite{Jakobsen:2021lvp}. Analogously, the scattering amplitudes approach has been employed to study radiation scattering off of a single massive source \cite{Cristofoli:2021vyo,Kol:2021jjc}, where a novel connection between  scattering amplitudes and black hole perturbation theory has emerged  \cite{Bautista:2021wfy}, shedding light on how to obtaining the higher-spin gravitational Compton amplitude \cite{BCGV} (see also \cite{Falkowski:2020aso,Chiodaroli:2021eug}). 

Even with the powerful scattering amplitudes techniques at hand, so far, radiative information from bodies moving on \textit{bounded} orbits has been obtained only via analytic continuation \cite{Kalin:2019inp, Kalin:2019rwq} of radiation observables of scattering bodies \cite{Herrmann:2021lqe,Dlapa:2021npj,Saketh:2021sri} (applying mainly in the large eccentricity limit). However, the almost $40$ year old derivation of the Einstein quadrupole formula from a Feynman diagrammatic perspective by Hari Dass and Soni \cite{HariDass:1980tq}, and the more recent derivation by Goldberger and Ridgway using the classical double copy \cite{Goldberger:2017vcg}, suggest that scattering amplitudes can indeed be used to derive gravitational radiation emitted from objects moving on general \textit{closed} orbits (including the zero eccentricity limit, i.e., quasi-circular orbits). In this work we follow this philosophy to compute the gravitational waveform emitted from an aligned spin BBH on general and quasi-circular orbits, up to quadratic order in the constituents spin at the leading order in the velocity expansion and to sub-leading order in the no-spin limit, from a 5-pt scattering amplitude of two massive particles exchanging and emitted gravitons. We contrast and compare these results to the analogous classical derivation of the corrections to the Einstein quadrupole formula using the well-established multipolar post-Minkowskian formalism  \cite{Thorne:1980ru, Blanchet:1985sp, Blanchet:1995fg, Blanchet:1998in, Blanchet:2013haa}.

We find perfect agreement between the classical and the scattering amplitudes derivation of all radiative observables we consider, to the respective orders in the spin and velocity expansions. Furthermore, we show that at leading order in the BBH velocities, there is a one-to-one correspondence between the BBH source's mass and current multipole moments, and the scalar and linear-in-spin 5-pt scattering amplitude, respectively. At quadratic order in the BHs' spin, we demonstrate explicitly that the corresponding contribution from the quadratic-in-spin scattering amplitude does not provide additional spin information at the level of the waveform; hence, we conjecture this to hold for higher-spin amplitudes as well, based on the aforementioned correspondence. Then, the leading in velocity, all orders-in-spin waveform, is obtained purely through the solutions to the equations of motion (EoM) of the conservative sector of the BBH. Furthermore, the gauge dependence of gravitational radiation information at future null infinity is a potential source of difficulty when comparing results obtained by different approaches. In this work, we provide evidence that gauge freedom partially manifests itself in the integration procedure appearing in the computation of the waveform directly from the scattering amplitude. For quasi-circular orbits, the orbit's kinematic variables are subject to certain relations, such that the gravitational waveform can take different forms without affecting the gauge invariant information contained in the total instantaneous gravitational wave energy flux.

This paper is organized as follows: In \cref{sec:classical_derivation}, we begin by reviewing the classical derivation of the conservative sector of the spinning BBH to all orders in the spins at leading PN. In \cref{sec:RadDyn}, we derive the associated gravitational wave emission from this system to all order in the BHs' spins. We then proceed with the scattering amplitudes derivation of the waveform in  \cref{sec:scatteringampl}, with the general formalism outlined in \cref{sec:general_approach_amp}. In \cref{sec:amplitude_double_copy} we present the relevant scattering amplitudes needed for the computation, obtained only through the Compton and the 3-pt amplitudes,  and subsequently use them  in \cref{sec:radiated_field_amplitudes} for   explicitly determining  the waveform. In section \cref{sec:results}, we briefly discuss the computation of the gauge invariant energy flux, and comment on the manifestation of the gauge freedom. We conclude with a discussion in  \cref{sec:discusion}. In this paper we use Greek letters $\alpha, \beta \dots$ for spacetime indices and Latin letters $i,j \dots$ for purely spatial indices. Furthermore, we use $G=c=1$ units throughout, assume $\varepsilon_{0123}=1$, and use the $2\nabla_{[\alpha}\nabla_{\beta]}\omega_\mu = R_{\alpha\beta\mu}{}^\nu \omega_\nu$ Riemann tensor sign convention. 

\section{Classical derivation}\label{sec:classical_derivation}

In order to approach the bound orbit from a classical point of view, we utilize an effective worldline action \cite{Porto:2005ac, Porto:2006bt, Porto:2008tb, Levi:2015msa, Levi:2014gsa, Porto:2016pyg, Levi:2018nxp}, parametrizing the complete set of spin-induced interactions of the two spinning BHs in the weak-field regime, at linear order in the gravitational constant, i.e. at PM order. As we are interested in \textit{bound}, as opposed to \textit{unbound}, orbits, we will be focusing on the leading PN contribution to the 1PM conservative sector at each order in the BHs' spins. In the following, we first briefly summarize the necessary conservative results established in Refs.~\cite{Levi:2016ofk, Tulczyjew:1959b, Barker:1970zr, Barker:1975ae, D'Eath:1975vw,Thorne:1984mz, Poisson:1997ha, Damour:2001tu,Hergt:2007ha, Hergt:2008jn, Levi:2014gsa, Vaidya:2014kza, Marsat:2014xea, Vines:2016qwa, Siemonsen:2017yux}. Using these results, we then tackle the radiative sector, utilizing the multipolar post-Minkowskian formalism \cite{Thorne:1980ru, Blanchet:1985sp, Blanchet:1995fg, Blanchet:1998in, Futamase:2007zz} (see also Ref.~\cite{Blanchet:2013haa} and references therein). We derive the transverse-traceless (TT) pieces of the linear metric perturbations, $h_{\mu\nu}^\text{TT}$, and the total instantaneous gravitational wave power, $\mathcal{F}$, radiated by this source to future null infinity. We achieve this, considering all orders in the spins, both for an aligned spin system on general orbits at leading order in velocities, as well as specialize to quasi-circular orbits at leading and first sub-leading orders in velocities. In this section, we work in the $-+++$ signature for the flat metric, and set $G=1$.


\subsection{Classical spinning Binary black hole} \label{sec:linKerr}

Let us begin by  briefly reviewing the approach to the conservative sector of the BBH dynamics at the respective orders in the weak-field and low-velocity regimes using an effective worldline action.  In sec.~\ref{sec:effBBHaction} we present the necessary spin-interactions to describe a rotating BH, while in sec.~\ref{sec:consvdyn} we review how an effective spinning BBH action, needed for the computation of the radiation field, can be derived.

\subsubsection{Effective binary black hole action} \label{sec:effBBHaction}

An effective description of a rotating black hole (BH), obeying the no-hair theorems, as a point particle with suitable multipolar structure in the weak-field regime rests solely on its worldline and spin degrees of freedom \cite{Porto:2016pyg, Porto:2006bt, Steinhoff:2014kwa, Marsat:2014xea, Levi:2015msa}. The former are given by a worldline $z^\mu(\lambda)$ of mass $m$, with 4-velocity  $u^\mu=dz^\mu/d\lambda$, while the latter are encoded in the BH's (mass-rescaled) angular momentum vector\footnote{This angular momentum vector $a^\mu =\varepsilon^\mu {}_{\nu \alpha\beta}u^\nu S^{\alpha\beta}/(2m)$ emerges from the spin tensor $S^{\alpha\beta}$ assuming the covariant spin supplementary condition, $p_\mu S^{\mu\nu}=0$, and a local body-fixed frame $e_A^\mu(\lambda)$. See, for instance, Ref.~\cite{Steinhoff:2014kwa} for details.} $a^\mu$, and local frame $e_A^\mu(\lambda)$. An effective worldline action, $S$, that entails the dynamics of such a BH (or, more generally, a compact object) in the weak-field regime was developed in Refs.~\cite{Porto:2005ac, Porto:2006bt, Porto:2008tb, Levi:2015msa, Levi:2014gsa}; see Ref.~\cite{Levi:2018nxp} for further details. This action $S[h,\mathcal{K}]$, describing a rotating compact object, is built considering all possible couplings of gravitational, $h=\{h_{\mu\nu}\}$, and object specific degrees of freedom, $\mathcal{K}=\{z^\mu,u^\mu,a^\mu,e^\mu_A\}$, requiring covariance, as well as reparameterization and parity invariance \cite{Goldberger:2009qd, Ross:2012fc, Levi:2015msa, Levi:2018nxp}. At  the 1PM level, a matching procedure between the linearized Kerr metric \cite{Harte:2016vwo, Vines:2017hyw, Vines:2016qwa} and the gravitational field $h_{\mu\nu}$, emanating from a generic compact object described by $S[h,\mathcal{K}]$, leads to a \textit{unique} set of non-minimal couplings between $h$ and $\mathcal{K}$. This ultimately results in an effective 1PM BH worldline action $S_\text{BH}[h,\mathcal{K}]$. This action can be extended to higher orders in $G$ in spins (see for instance Refs.~\cite{Levi:2020kvb, Levi:2020uwu, Levi:2020lfn,Siemonsen:2019dsu}).

It was shown in Ref.~\cite{Vines:2017hyw} that for a harmonic gauge linearized Kerr BH the infinite set of spin-couplings present in the 1PM effective worldline action $S_\text{BH}[h,\mathcal{K}]$ can be resumed into an exponential function. In a linear setup, a BH of mass $m$ traveling along the worldline $z^\mu(\lambda)$, sources the gravitational field, $g_{\mu\nu}=\eta_{\mu\nu}+h_{\mu\nu}^\text{Kerr}+\mathcal{O}(h^2)$, with \cite{Vines:2017hyw}
\begin{align}
h_{\mu\nu}^\text{Kerr}=4\mathcal{P}_{\mu\nu}{}^{\alpha\beta}\hat{\mathcal{T}}^\text{Kerr}_{\alpha\beta} \frac{1}{\hat{r}}, & & \hat{\mathcal{T}}_{\mu\nu}^\text{Kerr}=m \exp(a * \partial)_{(\mu}{}^\rho u_{\nu)}u_\rho.
\label{eq:LinKerr}
\end{align}
Here we define $(a * \partial)^\mu {}_\nu = \epsilon^\mu {}_{\nu\alpha\beta} a^\alpha \partial^\beta$ and introduced the trace reverser  $\mathcal{P}_{\mu\nu\alpha\beta}=(\eta_{\mu \alpha}\eta_{\nu \beta}+\eta_{\nu\alpha}\eta_{\mu\beta}-\eta_{\mu\nu}\eta_{\alpha\beta})/2$. Additionally, $\hat{r}$ labels the proper distance between the spacetime point $x$ and the worldline $z^\mu(\lambda)$, within the slice orthogonal to $u^\mu$ \cite{Vines:2017hyw}. In the following, we restrict ourselves to the leading PN part of the 1PM ansatz, since this is the natural setting for \textit{closed} orbits in the weak field regime. However, while we are expanding in $\epsilon_\text{PN}\sim v^2/c^2\sim GM/rc^2$, we consider all orders in the spins, i.e., consider $\epsilon_\text{spin}\sim \chi GM/rc^2$ non-perturbatively (here, $\chi$ the black hole's dimensionless spin parameter). To that end, we choose the Minkowski coordinate time $t$ to parameterize the worldline $z^\mu$, i.e., $\lambda\rightarrow t$, and expand the 4-velocity $u^\mu=(1,\boldsymbol{v})^\mu+\mathcal{O}(v^2)$, with $z^i=dz^i/dt=v^i$. Given this and utilizing the three-dimensional product $(a\times \partial)_i=\varepsilon_{ijk}a^j\partial^k$, the metric \eqref{eq:LinKerr} reduces to its leading PN form:
\begin{align}
\begin{aligned}
h^\text{Kerr}_{00} & \ =\left(2 \cosh (a\times \partial)-4 v_i \sinh (a\times \partial)^i\right)\frac{m}{\hat{r}} + \mathcal{O}(v^2),\\
h^\text{Kerr}_{0i} & \ =\left(4 v_i \cosh (a\times \partial)-2 \sinh (a\times \partial)_i\right)\frac{m}{\hat{r}} + \mathcal{O}(v^2),\\
h^\text{Kerr}_{ij} & \ =\left(2 \delta_{ij}\cosh (a\times \partial)-4 v_{(i}\sinh (a\times \partial)_{j)}\right)\frac{m}{\hat{r}} + \mathcal{O}(v^2).
\end{aligned}
\label{eq:hmunuPN}
\end{align}

Note that even at zeroth order in velocity, the solution contains non-trivial gravito-magnetic contributions, $h^\text{Kerr}_{0i}$, due to the presence of the BH spin. Conversely, an effective stress-energy distribution $T_{\mu\nu}$ can be derived that yields \eqref{eq:hmunuPN} via the linearized Einstein equations\footnote{At leading PN order, the spacetime effectively decomposes into space and time parts, yielding a simplification of the linearized Einstein equations: $\square_\text{ret.}^{-1}T_{\mu\nu}\rightarrow \Delta^{-1}T_{\mu\nu}$ (see Ref.~\cite{Blanchet:2013haa} for details).} $\square h_{\mu\nu}^\text{Kerr}=-16\pi \mathcal{P}_{\mu\nu}{}^{\alpha\beta}T_{\alpha\beta}$. This distribution has support only on the worldline $z^i(t)$ and, with the above parameterization, is given by
\begin{align}
T_{\mu\nu}(t,x^i) = \hat{\mathcal{T}}_{\mu\nu}^\text{Kerr}\delta^3(\boldsymbol{x}-\hat{\boldsymbol{z}}(t))+\mathcal{O}(\hat{v}^2).
\label{eq:KerrStressEn}
\end{align}

Collecting these within the worldline action, we can construct an effective binary BBH $S_\text{BBH}$ that encodes the conservative dynamics with the complete spin information at the leading PM level \cite{Vines:2017hyw} or leading PN level \cite{Vines:2016qwa, Siemonsen:2017yux}. That is, given two worldlines $z_{1,2}^\mu$, with velocities $u^\mu_{1,2}$, masses $m_{1,2}$, and two spin vectors $a^\mu_{1,2}$ -- conveniently collected in the sets $\mathcal{K}_{1,2}$ -- the spin interactions within the binary are obtained by integrating out the gravitational field in a Fokker-type approach \cite{Bernard:2015njp}. Following \cite{Vines:2017hyw, Siemonsen:2017yux}, in practice, the effective action for the second BH $S_\text{BH}[h,\mathcal{K}_2]$ (containing this BH's degrees of freedom $\mathcal{K}_2$) is evaluated at the metric of the first BH $h\rightarrow h_1$, such that $S_\text{BH}[h,\mathcal{K}_2]\rightarrow S_\text{BH}[h_1,\mathcal{K}_2]$. However, since the metric $h_1$, explicitly given in \eqref{eq:LinKerr}, is effectively a map from the gravitational degrees for freedom into that BHs' degrees of freedom, i.e., $h_{1}\rightarrow\mathcal{K}_{1}$, the BBH action $S_\text{BH}[h_1\rightarrow \mathcal{K}_1,\mathcal{K}_2]\rightarrow S_\text{BBH}[\mathcal{K}_1,\mathcal{K}_2]$, solely depends on the BHs' degrees of freedom.

\subsubsection{Conservative dynamics} \label{sec:consvdyn}

In order to write out the effective BBH action $S_\text{BBH}[\mathcal{K}_1,\mathcal{K}_2]$ explicitly, let us define the spatial separation $r^i=z^i_{1}-z^i_{2}$, with $r=|\boldsymbol{r}|$, between the two worldlines, as well as the spin sums $a^i_+=a_1^i+a_2^i$ and $a^i_-=a_1^i-a_2^i$. The angular velocity\footnote{The angular velocity tensor $\Omega^{\mu\nu}=e^\mu \cdot D e^\nu/d\lambda$ is defined by means of the body fixed frame $e_A^\mu(\lambda)$ along the worldline. The corresponding angular velocity vector is then given by $\Omega^i=\varepsilon^i{}_{jk}\Omega^{jk}/2$. See, for instance, Ref.~\cite{Steinhoff:2014kwa} for details.} 3-vectors $\Omega^i_{1,2}$ are introduced for completeness, however, the aligned-spin dynamics are independent of $\Omega_{1,2}^i$. Finally, we define the center of mass frame velocity $v^i=\dot{r}^i=v_1^i-v_2^i$. In Refs.~\cite{Vines:2016qwa, Siemonsen:2017yux} it was shown that after integrating out the gravitational degrees of freedom, as described in the previous section, the effective BBH action $S_\text{BBH}$ reduces to the two-body Lagrangian
\begin{align}
\begin{aligned}
\mathcal L_\text{BBH}= \bigg[\frac{m_1}{2}v_1^2 & \  + \frac{m_1}{2}\varepsilon_{ijk} a_1^i v^j_1
\dot{v}_1^k + m_1 a^i_1\Omega_{1,i} + (1\leftrightarrow 2) \bigg] \\
& \ + \bigg[ \cosh (a_+\times \partial)+ 2v_i\sinh (a_+\times \partial)^i \bigg]\frac{m_1 m_2}{r},
\end{aligned}
\label{eq:Leff}
\end{align}
at the leading PN level. Note that here and in the remainder of this section $\partial_i r^{-1}=\partial r^{-1}/\partial z^i_1=-\partial r^{-1}/\partial z_2^i$. So far, we have assumed a leading PN treatment at each order in spin, but kept the dynamics unrestricted. In the following we assume that the spin degrees of freedom are fixed, i.e., the spin vectors are independent of time, $\dot{a}_{1,2}=0$, and aligned with the orbital angular momentum of the system: $a_{1,2}^i\propto L^i$; hence, the motion is confined to the plane orthogonal to $L^i$. For later convenience, we define the unit vector $\ell^i$, such that $L^i=|\boldsymbol{L}|\ell^i$. Varying this action with respect to the worldline $z_1^i$, the classical EoM of the system are\footnote{The corresponding equation for $-\dot{v}^i_2$ emerges from the right hand side of \eqref{eq:EoM} under the replacement $a_1^i\leftrightarrow a_2^i$.} \cite{Siemonsen:2017yux, Vines:2016qwa}
\begin{align}
\begin{aligned}
\dot{v}_1^i= \Big(\partial^i-  & \ \varepsilon^i{}_{jk}a_1^kv^l\partial_l\partial^j\Big)\cosh(a_+\times\partial)\frac{ m_2}{r}\\
& \ +2 \left( v_j\partial^i-\delta^i_jv^k\partial_k\right)\sinh(a_+\times \partial)^j\frac{ m_2}{r} + \mathcal{O}(v^2).
\end{aligned}
\label{eq:EoM}
\end{align}
A geometric approach using oblate spheroidal coordinates \cite{Vines:2016qwa, Harte:2016vwo, Vines:2017hyw} or an algebraic approach, exploiting properties of the Legendre polynomials \cite{Siemonsen:2017yux}, under the assumption that the motion takes place in the plane orthogonal to the spin vectors, can be used to resum the series of differential operators in \eqref{eq:EoM}.

In order to present the contribution of the conservative sector needed for the radiative dynamics, we specialize to the center of mass frame for the rest of this section. The transformation into the center of mass variables $r^i$ based on \eqref{eq:Leff} (and using the total mass $M=m_1+m_2$), is corrected by the presence of the spins only at sub-leading orders in velocities:
\begin{align}
\begin{aligned}
\begin{split}
z^i_1=\frac{m_2}{M}r^i-b^i, \qquad z^i_2=-\frac{m_1}{M}r^i-b^i, \quad b^i:=\frac{1}{M}\varepsilon^i{}_{jk}(v^j_1 S^k_1+v^j_2 S^k_2).
\end{split}
\label{COMframe}
\end{aligned}
\end{align}
In this center of mass frame, the EoM are readily solved for quasi-circular motion. In that scenario, the separation $r^i$ is related to its acceleration $\ddot{r}^i$ by $r^i=-\ddot{r}^i/\omega^2$, where $\omega$ is the system's orbital frequency. This ansatz picks out the quasi-circular orbits allowed by the BBH EoM \eqref{eq:EoM} and is equivalent to finding a relation between the frequency $x=(M \omega)^{2/3}$, the BHs spins $a^i_{1,2}$, and the separation of the binary $r$. This relation, at the leading PN order at each order in the BHs' spins, is given by \cite{Siemonsen:2017yux}
\begin{align}
r(x)=\sqrt{\frac{M^2}{x^2}+\bar{a}_+^2}\left(1-\frac{x^{3/2}M}{3}\frac{\bar{\sigma}^*+2\bar{a}_+}{M^2+x^2\bar{a}_+^2}\right),
\label{eq:consvsol}
\end{align}
where $\bar{\sigma}^*=(m_2 \bar{a}_1+m_1 \bar{a}_2)/M$ and we defined $\bar{a}_{1,2}=a_{1,2}^i\ell_i$. It should be emphasized that the even-in-spin part of \eqref{eq:consvsol} contains only $\mathcal{O}(v^0)$ information, while the odd-in-spin pieces are non-zero only at first sub-leading order in velocities, at $\mathcal{O}(v^1)$. This solution can then be used to compute gauge invariant quantities of the conservative sector, such as the total binding energy and angular momentum \cite{Siemonsen:2017yux}.

\subsection{Linearized metric perturbations at null infinity} \label{sec:RadDyn}

With the conservative results in hand, in this subsection, we compute the gravitational waves from the BBH system at future null infinity. In sec.~\ref{sec:psi4generalapproach} we briefly review the general approach of mapping the source's multipole moments into the radiation field, while in sec.~\ref{sec:hijclassical} we derive the TT part of the linear metric perturbations (the gravitational waves) at null infinity utilizing this mapping. 

\subsubsection{General approach} \label{sec:psi4generalapproach}

A natural choice of gauge invariant quantity capturing the radiative dynamics at null infinity is the Newman-Penrose Weyl scalar $\Psi_4$. This contains both polarization states, $h_+$ and $h_\times$, of the emitted waves, which are the observables measured by gravitational wave detectors. Upon choosing a suitable null tetrad, the TT part of the gravitational field, $h_{\mu\nu}^\text{TT}$, can be related to $\Psi_4$:
\begin{align}\label{eq:NP scalar}
\Psi_4\sim\ddot{h}_+-i\ddot{h}_\times=\bar{m}^\mu \bar{m}^\nu \ddot{h}^\text{TT}_{\mu\nu}.
\end{align}
The complex conjugate pair $\{m^\alpha, \bar{m}^\alpha\}$ is typically defined with respect to the flat spherically symmetric angular coordinate directions $m =(\boldsymbol{\Theta}+i \boldsymbol{\Phi})/\sqrt{2}$. With this choice in place, we restrict our attention to the spatial components $h^\text{TT}_{ij}$, as these contain the full information of $\Psi_4$, i.e., the radiative, non-stationary, degrees of freedom\footnote{As we will see below, this choice of purely spatial $m^\alpha$ is equivalent to choosing a gauge, in which the graviton polarization tensor is also purely spatial.}. 

In the previous section, we summarized the leading PN conservative dynamics of a spinning BBH to all orders in their spins. Given this, the well-established multipolar post-Minkowskian formalism \cite{Thorne:1980ru, Blanchet:1985sp, Blanchet:1995fg, Blanchet:1998in, Blanchet:2013haa} is ideally suited to determine the time-dependent metric perturbations at null infinity. Within this framework, the stress energy distribution of the source, $T_{\mu\nu}^\text{source}$, is mapped into a set of mass and current symmetric and trace free (STF) source multipole moments $\mathcal{I}_{i_1\dots i_\ell}(t)$ and $\mathcal{J}_{i_1\dots i_\ell}(t)$. We denote $\langle i_1\dots i_\ell\rangle$ as the STF projections  of the indices $i_1\dots i_\ell$. Then the STF multipole moments evaluated at the retarded time $T_R=t-R$ are defined by \cite{Blanchet:2013haa}
\begin{align}
\begin{aligned}
\mathcal{I}_{i_1\dots i_\ell}= & \ \int d\mu\left(\delta_\ell x_{\langle i_1\dots i_\ell\rangle} \Sigma - f_{1,\ell}\delta_{\ell+1}x_{\langle i i_1\dots i_\ell\rangle} \dot{\Sigma}^i + f_{2,\ell}\delta_{\ell+2}x_{\langle ij i_1\dots i_\ell\rangle}\ddot{\Sigma}^{ij} \right)(\boldsymbol{x},T_R+zr),\\
\mathcal{J}_{i_1\dots i_\ell}= & \ \int d\mu \ \varepsilon_{ab\langle i_\ell}\left(\delta_\ell x_{i_1\dots i_{\ell-1}\rangle}{}^a\Sigma^b- g_{1,\ell}\delta_{\ell+1} x_{i_1\dots i_{\ell-1}\rangle c}{}^a \dot{\Sigma}^{bc} \right)(\boldsymbol{x},T_R+zr),
\label{eq:generalsourcemultipoles}
\end{aligned}
\end{align}
where $x_{i_1\dots i_\ell}=x_{i_1}\dots x_{i_\ell}$,
\begin{align}
f_{1,\ell}=\frac{4(2\ell+1)}{(\ell+1)(2\ell+3)}, & & 
f_{2,\ell}=\frac{2(2\ell+1)}{(\ell+1)(\ell+2)(2\ell+5)}, & & g_{1,\ell}=\frac{2\ell+1}{(\ell+2)(2\ell+3)},
\end{align}
and the integration measure $\int d\mu=\text{FP}\int d^3 \boldsymbol{x} \int_{-1}^1 dz$. The source energy-momentum distribution enters in $\Sigma$, via (valid only at leading PN orders)\footnote{At sub-leading PN orders, the stress energy of the emitted gravitational waves contributes to $\Sigma$.}
\begin{align}
\Sigma=T^{00}+T^{ij}\delta_{ij}, & & \Sigma^i = T^{0i}, & & \Sigma^{ij}=T^{ij}.
\end{align}
The source' finite size retardation effects are contained in the $z$-integral with $\delta_\ell=\delta_\ell(z)$ in \eqref{eq:generalsourcemultipoles}, which are given explicitly in eq.~(120) of Ref.~\cite{Blanchet:2013haa}. At the orders considered in this work, at the leading PN orders, finite size-retardation effects vanish and the $z$-integral trivializes: $\int_{-1}^1 dz \  \delta_\ell(z)f(\boldsymbol{x},T_R+rz)=f(\boldsymbol{x},T_R)+\mathcal{O}(v^2)$. We discuss in \cref{sec:general_approach_amp}, how a similar structure as in \eqref{eq:generalsourcemultipoles} appears in the scattering amplitudes approach, as well as what precisely encapsulates the ``finite size" retardation effects in that context. The lowest order moments $\mathcal{I}$, $\mathcal{I}_i$, and $\mathcal{J}_i$ are constants of motion representing the total conserved energy, center of mass position and total angular momentum, respectively. Only for $\ell\geq 2$, do the multipoles contribute non-trivially.

A matching scheme enables to directly relate these functionals for the source's stress-energy distribution, to the radiation field at null infinity (at 1PM order)\footnote{Beyond linear theory, corrections to these multipole moments are necessary \cite{Blanchet:2013haa}.} \cite{Blanchet:2013haa}
\begin{align}
h_{ij}=-4 \sum_{\ell= 2}^\infty \frac{(-1)^\ell}{\ell!}\left[\partial_{i_1\dots i_{\ell-2}}\ddot{\mathcal{I}}_{ij}{}^{i_1\dots i_{\ell-2}}R^{-1}+\frac{2\ell}{\ell+1}\partial_{a i_1 \dots i_{\ell-2}}\varepsilon^{ab}{}_{(i}\dot{\mathcal{J}}_{j)b}{}^{i_1\dots i_{\ell-2}}R^{-1} \right].
\label{eq:generalhij}
\end{align}
 Here $\partial_a R^{-1}=-N_a/R^2$ is to be understood as the derivative in the background Minkowski spacetime, where $N_a$ is radially outwards pointing from the source to spatial infinity, with $N_a N^a=1$. To solely focus on the radiation at null infinity, we work to leading order in the expansion in $R^{-1}$. Therefore, the spatial derivatives in \eqref{eq:generalhij} act purely on the source multipole moments, and there, can be traded for time derivatives: $\partial_a f(t-R)=-\dot{f}N_a$. Similarly, the total instantaneous gravitational wave energy flux $\mathcal{F}$ can be derived directly from the source multipole moments \cite{Blanchet:2013haa}.

\subsubsection{Gravitational radiation from spinning binary black hole} \label{sec:hijclassical}

At the 1PM level, non-linear effects vanish such that the energy-momentum of the BBH is simply the superposition of two linearized Kerr BHs' energy momentum distributions \eqref{eq:KerrStressEn},  $T_{\mu\nu}^\text{source}=T_{\mu\nu}^{\text{Kerr},1}+T_{\mu\nu}^{\text{Kerr},2}$. This superposition holds in the conservative sector, while the radiative dynamics are derived directly from derivatives acting on $T_{\mu\nu}^\text{source}$ in the manner described in the previous section. From the scattering amplitudes perspective, this superposition is reflected in the \textit{only }two channel factorization of the  classical 5-pt amplitude,  into the product of a 3-pt amplitude and the gravitational Compton amplitude, as we will see in \cref{sec:amplitude_double_copy}. 

\textit{Leading order in velocities} -- As the radiative quantities $h_{ij}^\text{TT}$ and $\mathcal{F}$ depend on time derivatives of the source multipole moments, we focus on time-dependent terms after fixing the angular momentum dynamics. For the case of the above spinning BBH with aligned spins, at the leading PN order, we expand the source $T_{\mu\nu}^\text{source}$ analogously to \eqref{eq:hmunuPN}. Given this, the resulting leading-in-velocity contributions to the source multipole moments, utilizing \eqref{eq:generalsourcemultipoles}, are \cite{Buonanno:2012rv, Marsat:2014xea, Siemonsen:2017yux}
\begin{align}
\mathcal{I}^{ij}_{(0)} = m_1 z_1^{\langle ij\rangle}+(1\leftrightarrow 2), & & \mathcal{J}^{ij}_{(0)} = \frac{3}{2}S_1^{\langle i}z_1^{j\rangle} +(1\leftrightarrow 2),
\label{eq:multipolemomentv0}
\end{align}
where $(0)$ indicates the order in velocities. It should be stressed that these are \textit{all} multipoles needed for the gravitational waveform to \textit{all} orders in the BHs' spins, at leading order in velocity \cite{Siemonsen:2017yux}. From the amplitudes perspective, this will be reflected in the need for only the scalar and linear-in-spin scattering amplitudes at the leading orders in velocities. While all higher-order spin terms in the source multipole moments vanish identically, spin contributions to the waveform at arbitrary order in the spin expansion could enter through the solution to the EoM \eqref{eq:consvsol}. We see below that this solution to the classical EoM \eqref{eq:EoM} introduces non-zero contributions at arbitrary orders in the BHs' spins for quasi-circular orbits.

Given \eqref{eq:multipolemomentv0}, the metric perturbation at null infinity, for general orbits at zeroth order in velocities, assuming aligned spins, is
\begin{align}
h^{(0)ij}_{S^\infty}(T_R,R,\boldsymbol{N},\boldsymbol{z}_1,\boldsymbol{z}_2) = \frac{2 m_1}{R}\bigg\{\frac{d^2}{dt^2}\left[z_1^i z_1^j\right] + \varepsilon_{pq}{}^{(i}\left(a_1^{j)}\dot{v}_1^p+\dot{v}_1^{j)}a_1^p\right)N^q\bigg\}\bigg|_{t=T_R}+(1\leftrightarrow 2),
\label{eq:classicalhijgeneralorbit}
\end{align}
i.e. the Einstein Quadrupole formula with spinning corrections for a binary system. We specialize to quasi-circular orbits by introducing the orthogonal unit vectors
\begin{align}
n^i=r^i/r = (\cos \omega t, \sin\omega t,0)^i, & & \lambda^i=v^i/v=(-\sin\omega t, \cos\omega t, 0)^i,
\label{eq:circorbits}
\end{align}
in the center of mass frame that rotate with frequency $\omega$ in the orbital plane. The spin vectors $a_{1,2}^i\propto \ell^i$ are aligned orthogonal to the orbital plane, $n^i\lambda^j \varepsilon_{ij}{}^k=\ell^k$, such that $\ell^i=(0,0,1)^i$. Furthermore, the TT projector
\begin{align}
\Pi^{ij}{}_{kl}=P^i{}_k P^j{}_l-\frac{1}{2}P^{ij}P_{kl}
\label{eq:TTprojector}
\end{align}
is defined relative to $N_a$, where $P_{ij}=\delta_{ij}-N_i N_j$. Utilizing \eqref{COMframe}, together with the solution \eqref{eq:consvsol} to the EoM, as well as \eqref{eq:multipolemomentv0}, the gravitational waves emitted by the spinning BBH to all orders in the BHs' spins is conveniently written as
\begin{align}
h^{\text{TT}}_{ij}(T_R) =\frac{2\mu}{R} \ \Pi_{ij}{}^{ab} \hat{h}_{ab}\Big|_{t=T_R},
\label{eq:hijeasier}
\end{align}
where at leading order in velocities, we have $\hat{h}^{(0)}_{ab}=\hat{h}^{(0),\text{I2}}_{ab}+\hat{h}^{(0),\text{J2}}_{ab}$, with
\begin{align}
\label{eq:hijcircularv0}
\begin{aligned}
\hat{h}^{(0),\text{I2}}_{ab} = & \ -2x \left(1+\frac{\bar{a}_+^2x^2}{M^2}\right)(n_an_b-\lambda_a\lambda_b) \\
\hat{h}^{(0),\text{J2}}_{ab} = & \ -\frac{x^2 \bar{a}_-}{M}\sqrt{1+\frac{\bar{a}_+^2x^2}{M^2}},\ \varepsilon_{kl(a}(\ell_{b)}n^k+n_{b)}\ell^k)N^l.
\end{aligned}
\end{align}

Notice that here, the odd-in-spin contribution, $\hat{h}^{(0),\text{J2}}_{ab}$, is a series that has non-zero coefficients at arbitrary orders in spin, arising from the odd part of the solution  \eqref{eq:consvsol},  while, on the other hand, the even-in-spin part, $\hat{h}^{(0),\text{I2}}_{ab}$, provides coefficients that vanish for $\mathcal{O}(a^{\ell\geq 3})$. This is analogous to the cancellations observed in the conservative and radiative sectors reported in Ref.~\cite{Siemonsen:2017yux}. We find agreement with  the results reported in Refs.~\cite{Kidder:1992fr, Kidder:1995zr, Buonanno:2012rv} to the respective finite order in spin.  To check for consistency to all orders in spin, the gravitational wave modes are extracted from the spatial part of the metric perturbations, in \eqref{eq:hijcircularv0}, by projecting onto a suitably defined basis of spin-weighted spherical harmonics, ${}_{-2}Y_{\ell m}(\Theta, \Phi)$. Explicitly, the gravitational wave modes $h^{\ell m}$ are defined to be $h^{\ell m}=\int d\Omega \ {}_{-2}\bar{Y}_{\ell m}(\Theta, \Phi) \bar{m}^\mu \bar{m}^\nu h^\text{TT}_{\mu\nu}$. These modes, obtained from \eqref{eq:hijcircularv0} in conjunction with the above defined polarization tensor $\bar{m}^\alpha\bar{m}^\beta$, agree with the results in Ref.~\cite{Siemonsen:2017yux} to all orders in the BHs' spins at leading order in their velocities.

\textit{Sub-leading order in velocities} -- The sub-leading corrections to the above radiation field are obtained in much the same way. The additional contributions to the source multipole moments, beyond the leading pieces \eqref{eq:multipolemomentv0}, at sub-leading orders in velocities are \cite{Buonanno:2012rv, Marsat:2014xea, Siemonsen:2017yux}
\begin{align}
\begin{aligned}
\mathcal{I}^{ij}_{(1)} = & \ \frac{4}{3}\left(2v_1^a S^b_1\varepsilon_{ab}{}^{\langle i}z_1^{j \rangle}-z_1^aS_1^b\varepsilon_{ab}{}^{\langle i} v^{j\rangle} \right) +(1\leftrightarrow 2), \\
\mathcal{I}^{ijk}_{(1)} = & \ m_1 z_1^{\langle ijk\rangle}-\frac{3}{m_1}S_1^{\langle i}S_1^j z_1^{k\rangle} +(1\leftrightarrow 2), \\
\mathcal{J}^{ij}_{(1)} = & \ m_1 z_1^av_1^b\varepsilon_{ab}{}^{\langle i}z_1^{j\rangle} + \frac{1}{m_1}v_1^a S_1^b\varepsilon_{ab}{}^{\langle i}S_1^{j\rangle}+(1\leftrightarrow 2), \\
\mathcal{J}^{ijk}_{(1)} = & \ 2S_1^{\langle i}z_1^{jk\rangle}+(1\leftrightarrow 2).
\label{eq:multipolemomentv1}
\end{aligned}
\end{align}
Also here, we focused only on those pieces that are time-dependent, i.e., that will contribute non-vanishing terms in $h_{ij}^{(1)\text{TT}}$. Additionally, as pointed out above, these are \textit{all} necessary contributions for the full all orders-in-spin information at sub-leading orders in velocities (at leading PN order) \cite{Siemonsen:2017yux}. Using this, together with the mapping \eqref{eq:generalhij}, the decomposition \eqref{eq:hijeasier}, and $\hat{h}^{(1)}_{ab}=\hat{h}^{(1),\text{I2}}_{ab}+\hat{h}^{(1),\text{J2}}_{ab}+\hat{h}^{(1),\text{I3}}_{ab}+\hat{h}^{(1),\text{J3}}_{ab}$, the sub-leading contribution $h_{ij}^{(1)\text{TT}}$ to all orders in spin from a spinning binary black hole on quasi-circular orbits are
\begin{align}
\label{eq:hijcircularv1}
\begin{aligned}
\hat{h}^{(1),\text{I2}}_{ab}= & \   \frac{4 x^{5/2}}{3M^3}\left(2\bar{a}_+M^2+\left(M^2-2 r_e^2 x^2\right)\bar{\sigma}^*\right)\left(n_a n_b-\lambda_a \lambda_b\right), \\
\hat{h}^{(1),\text{J2}}_{ab}= & \  \frac{ x^{5/2}}{3M^4 r_e}\left[2r_e^4 x^2 \delta m+\bar{a}_- M\left(2\bar{a}_+(M^2-r_e^2x^2)+3r_e^2 x^2 \bar{\sigma}+M^2\bar{\sigma}^*\right)\right] \\ 
& \ \times\varepsilon_{pq(a}N^q \left(n^p\ell_{b)}+n_{b)}\ell^p\right),\\
\hat{h}^{(1),\text{I3}}_{ab}= & \  \frac{r_e x^{9/2}}{15M^4}\left[15 \bar{a}_+ \bar{a}_- M \ell_{\langle a}\ell_b \lambda_{k\rangle}-r_e^2\delta m\left(30 \lambda_{\langle a}\lambda_b\lambda_{k\rangle}-105 n_{\langle a}n_b \lambda_{k\rangle}\right)\right]N^k, \\
\hat{h}^{(1),\text{J3}}_{ab}= & \ -\frac{48 r_e^2 x^{9/2} \bar{\sigma}^*}{6M^3}\epsilon_{pq(a} \delta_{b)k} n^{(k}\lambda^p\ell^{e)}N^q N_e.
\end{aligned}
\end{align}
Here $r_e=(\bar{a}_+^2+M^2/x^2)^{1/2}$, which is just the leading-in-velocities (even-in-spin) solution to the classical EoM \eqref{eq:consvsol} for quasi-circular orbits. We check the gravitational wave modes obtained from \eqref{eq:hijcircularv1} with those presented in Ref.~\cite{Siemonsen:2017yux} and find agreement to all orders in spin. Additionally, we compute the gauge invariant gravitational wave energy flux with the above result together with the leading-in-velocities radiation field and find agreement with results reported in \cite{Marsat:2014xea, Siemonsen:2017yux} (see also a detailed discussion in sec.~\ref{sec:results} below). Finally, in order to compare to the scalar amplitude at first sub-leading orders in the BHs' velocities in \cref{sec:subleading-v}, we also present the radiation field of a non-spinning BBH system on general orbits, to sub-leading order in velocities:
\begin{align}\label{eq:h1_v1_scalar}
h^{(1),ij}_{S^0,\text{TT}}=\frac{2m_1}{3R}\Pi^{ij}{}_{ab}\left[4\varepsilon_{pq}{}^{(a}\Big\{ \partial_t^2 (\varepsilon_{cd}{}^{e)}z_1^c v_1^d\delta_e{}^{\langle b}z_1^{p\rangle})\Big\}N^q +N_k\partial_t^3 (z_1^{\langle a}z_1^b z_1^{k\rangle})\right]+(1\leftrightarrow 2).
\end{align}

\section{Scattering Amplitudes derivation} \label{sec:scatteringampl}

In the previous sections, we obtained the form of the  gravitational waves emitted from a spinning BBH on general \textit{closed} orbits with aligned spins, to leading order in the BHs'  velocities  \eqref{eq:classicalhijgeneralorbit}   [and on quasi-circular orbits given in \eqref{eq:hijcircularv0}], whereas  at sub-leading  order in $v$,  and for quasi-circular orbits, we derived  \eqref{eq:hijcircularv1}, at each order (and to \textit{all} orders) in the BHs' spins. In the following, we show that these results follow directly from the classical limit of a 5-pt spinning \textit{scattering} amplitude.
More precisely, at leading order in velocity there is a one-to-one correspondence between the source's mass and current  multipole moments \eqref{eq:multipolemomentv0}, and the scalar and linear-in-spin  contribution to the scattering amplitude, respectively. This correspondence allow us to derive the linear in spin, general orbit result for the radiated gravitational field \eqref{eq:classicalhijgeneralorbit}, from an amplitudes perspective. At  quadratic order in the BHs' spins, and for quasi-circular orbits, we demonstrate that the contribution from the quadratic in spin amplitude is canceled by the contribution of the  scalar amplitude in conjunction with the $\mc{O}(S^2)$-piece of the EoM \eqref{eq:EoM}. This leaves only the quadrupole field, \eqref{eq:einstein-quadrupole}, supplemented with the solution to the EoM \eqref{eq:consvsol}, to enter at quadratic order in spin. Although we explicitly demonstrate the cancellation for quasi-circular orbits and up to quadratic order in spin only, we expect this theme to continue to hold for more complicated bound orbits, as well as to higher spin orders in the 5-pt scattering amplitude, as suggested by the classical multipole moments \eqref{eq:multipolemomentv0}. At sub-leading orders in the BHs' velocities the situation becomes more complicated; there, we demonstrate the matching of the amplitudes to the classical computation in the spin-less limit for quasi-circular orbits, and briefly comment on extensions to higher orders in spin. In this section, we use the mostly minus signature convention for the flat metric $\eta_{\mu\nu}=\rm{diag}(1,-1,-1,-1)$.\\
\begin{figure}[h]
\centering
\includegraphics[width=0.5\textwidth]{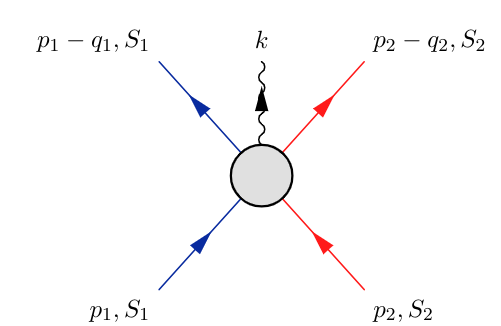}
\caption{Bremsstrahlung (outgoing graviton) emitted during the scattering of two spinning massive bodies (massive spin-$s$ quantum particles) exchanging gravitational/electromagnetic waves.}
\label{fig:5pt-amplitude}
\end{figure}

\subsection{General approach}\label{sec:general_approach_amp}

To compute the radiated field at future null infinity from the BBH system we follow the approach used by Goldberger and Ridgway in  \cite{Goldberger:2017vcg} to derive the Quadrupole formula, and extend it to include relativistic and spin effects. This approach is  based on the classical EoM for the orbiting objects in combination with the corresponding 5-pt (spinning) scattering amplitude (see Figure \ref{fig:5pt-amplitude}). It is valid for BBHs whose components have Schwarzschild radii $r_{1,2}=2m_{1,2}$ much smaller than their spatial separation $r$, i.e., $r_{1,2}\ll r$, while the radiation field wavelength is much bigger than the size of the individual components $\lambda\gg m_{1,2}$, as well as the size of the system $\lambda\gg r$\footnote{In the long distance separation regime, radiation reaction effects can be neglected, since they become important only when the separation of the two bodies is comparable to the system's gravitational radius \cite{Misner1973} eq. $36.11$.}.  Therefore, we expect our results to be situated in the PN regime of the binary inspiral\footnote{We stress that even though we concentrate mostly in the computation of gravitational waveform, an analogous derivation follows for  electromagnetic radiation, as   already pointed out in \cite{Goldberger:2017vcg}.}.

Let us start by noting that in the limit in which $R\rightarrow\infty$, where $R$ is the distance from the source to the observer (i.e., the radial coordinate in Bondi-Sachs gauge) as defined above, the time domain waveform  at retarded time $T_{R}$, has the asymptotic form \cite{Maggiore:1900zz} 
\begin{equation}\label{eq:radiated field}
h_\text{TT}^{ij}(T_{R},R,\boldsymbol{N},\boldsymbol{z}_1,\boldsymbol{z}_2)=\frac{\kappa}{16\pi R }\Pi^{ij}{}_{ab}\int d\bar{\omega}e^{-i\bar{\omega}\,T_{R}}T^{ab}(\bar{\omega},\boldsymbol{N},\boldsymbol{z}_1,\boldsymbol{z}_2).
\end{equation}
Here $\kappa^2=32\pi$ (recall we set $G=1$),  $\bar{\omega}$ is the frequency of the radiated wave with four  momentum $k^\mu=\bar{\omega}N^\mu=\bar{\omega}(1,\boldsymbol{N})^\mu$, and $\Pi^{ij}{}_{ab}$ is the TT-projector defined in \eqref{eq:TTprojector}.
As above, the locations of the binary's components are denoted by $z^i_{1,2}$. Analogous to the previous section,  we focus only  on the spatial components of $h^{\mu\nu}$, which contain all the radiative degrees of freedom. In what follows we also simplify the notation for the source $T^{ab}(\bar{\omega},\boldsymbol{N},\boldsymbol{z}_1,\boldsymbol{z}_2)\to T^{ab}(k,\boldsymbol{z}_1,\boldsymbol{z}_2) $, where it is understood that $k^\mu$ has implicit the dependence in both, $\bar{\omega}$ and $\boldsymbol{N}$. 

The source $T^{ab}(k,\boldsymbol{z}_1,\boldsymbol{z}_2)$, is related directly to the 5-pt scattering amplitude in Figure \ref{fig:5pt-amplitude}; therefore, in order to focus on the spatial components, it is sufficient to work in a gauge, in  which the graviton polarization tensor $\epsilon^{\mu\nu} =\epsilon^\mu\epsilon^\nu$, is the tensor product of two purely spatial  polarization vectors $\epsilon^\nu$. From the classical perspective, this choice of gauge is analogous to the conjugate pair $\{m^\alpha,\bar{m}^\alpha\}$ (defined in \cref{sec:psi4generalapproach}) to be purely spatial. Notice, however, the radiation field computed from a 5-pt scattering amplitude, and the corresponding field computed classically in the previous section, can in general differ by a gauge transformation. As shown below, this is directly related to a freedom in choice of an integration by parts (IBP) prescription in \eqref{eq:radiated field}.

We proceed by writing the explicit form of the source $T^{ij}(k,\boldsymbol{z}_1,\boldsymbol{z}_2) $ in terms of the  classical 5-pt scattering amplitude. In the classical computation, $T^{ij}$ corresponds to the source entering on the right hand side of  field equations, at a given order in perturbation theory. To leading order, for scalar particles, it was shown in \cite{Goldberger:2017vcg}  that the source can be rearranged in such a way, so that the scalar  5-pt   amplitude can be identified as the main kinematic object entering the graviton phase space integration, as well as the integration over the  particles proper times (which account for the particles history). In this paper we propose that formula to also hold for spinning particles.  That is,
\begin{equation}\label{eq:source_some1}
T^{ij}(k,\boldsymbol{z}_1,\boldsymbol{z}_2) = \frac{i}{m_1 m_2}\int d\tau_{1}d\tau_{2}\hat{d}^{4}q_{1}\hat{d}^{4}q_{2}\hat{\delta}^4\left(k-q_{1}-q_{2}\right)e^{iq_{1}{\cdot}z_{1}}e^{iq_{2}{\cdot}z_{2}}\langle M_{5}^{ij}(q_{1},q_{2},k)\rangle.
\end{equation}
Here $\langle M_{5}^{ij}\rangle$ is the classical 5-pt amplitude \footnote{\textcolor{black}{To motivate this formula, although this is by no means a formal derivation,  as already observed in \cite{Goldberger:2017vcg} for the tree-level amplitude, we can take the expression for the  radiation kernel  eq. $(4.42)$  in the KMOC original work  \cite{Kosower:2018adc} 
\begin{equation}
    \mathcal{J}=\lim_{\hbar\to 0}\frac{1}{m_1m_2}\Big\langle
    \int \prod_{i=1}^2\left[
    \hat{d}^4q_i \hat{\delta}(v_i{\cdot}q_i-q_i^2/(2m_i))e^{ib_i{\cdot}q_i} \right]\hat{\delta}^4(k-q_1-q_2)M_5
   \Big\rangle,
\end{equation}
and use the integral representation for the on-shell delta functions $\delta(x)\sim\int dy e^{ix y}$. Identifying the   asymptotic trajectories for the particles $z_i(\tau_i) = b_i^\mu +v_i^\mu \tau_i$, plus a quantum correction  $z_{Q}^\mu(\tau_i) = -\frac{ q_i}{2m_i}\tau_i $, and upon restoring the $\hbar$-counting in the exponential,   the radiation kernel can be rewritten as 
\begin{equation}\label{eq:radiation_kernell_gen}
    \mathcal{J}=\lim_{\hbar\to 0}\frac{1}{m_1m_2}\Big\langle
    \int \prod_{i=1}^2\left[
    d\tau_i\hat{d}^4q_i  e^{i q_i{\cdot}(z_i(\tau_i)+  \hbar z_{Q}(\tau_i))} \right]\hat{\delta}^4(k-q_1-q_2)M_5
   \Big\rangle.
\end{equation}
In the  classical limit, and to leading order in perturbation theory,   we can simply drop quantum correction to the particles trajectories    $z_{Q}^\mu(\tau_i)  $, and recover the formula \eqref{eq:source_some1} upon  promoting   $z_i(\tau_i)$ to be valid for generic time dependent  orbits. 
}} . Conventions for the particles' momenta  and the spins are shown in  Figure \ref{fig:5pt-amplitude}, with the condition for  momentum conservation     $q_{1}+q_{2}=k$. We have used the notation $\hat{d}^4q_i = \frac{d^4q_i}{(2\pi )^4} $, and similarly for the momentum-conserving delta function  $\hat{\delta}^4(p) = (2\pi)^4\delta^4(p)$. The position vectors are $z^\mu_A= (\tau_A,\boldsymbol{z}_A)^\mu$, with $A=1,2$, as described in \cref{sec:effBBHaction}, where the proper times $\tau_A$, parametrize the BHs' trajectories. Here the product of the exponential functions,  $\prod_Ae^{i q_A\cdot z_A}$, represents the two-particles initial state where each  particle is taken to be in a plane-wave state. This is nothing but the Born approximation in Quantum Mechanics (See also \cite{HariDass:1980tq}). 

We have striped away the graviton polarization tensor in \eqref{eq:source_some1}, assuming there exist the aforementioned gauge fixing for which the graviton polarization tensor is purely spatial. We can further  rewrite the source  using the  symmetric variable $q=(q_{1}-q_{2})/2$, as well as exploiting the  momentum conserving delta function to remove one of the $q_{i}$-integrals. The result reduces to
\begin{equation}\label{eq:source_some2}
T^{ij}(k,\boldsymbol{z}_1,\boldsymbol{z}_2)=\frac{i}{m_1 m_2}\int d\tau_{1}d\tau_{2}\hat{d}^{4}qe^{ik \cdot\tilde{z}}e^{-iq\cdot z_{21}}\langle M_{5}^{ij}(q,k)\rangle,
\end{equation}
where $\tilde{z}=(z_{1}+z_{2})/2$ and $z_{BA}=z_{B}-z_{A}$. Since we are interested in the bound-orbit problem, we take the slow-motion limit. Therefore, we can write the momenta of the BHs moving on \textit{closed}  orbits in the form $p_{1,2}^{\mu}=m_{1,2}v^{\mu}_{1,2}$. As noted above, we choose the frame in which $v_{1,2}^{\mu}=(1,\boldsymbol{v}_{1,2})^\mu+\mathcal{O}(v_{1,2}^{2})$, where $v^i_{1,2}=  dz^i_{1,2}/dt$, i.e. with the   proper times $\tau_{1,2}$  replaced by the coordinate time (see details below). On the other hand, in the closed orbits scenario the typical frequency of the orbit $\omega$, scales with $v$ as $\omega\sim v/r$, where $\omega=v/r$ for quasi-circular orbits (see also \eqref{eq:circorbits}). In this bound-orbits case, the integration in $q$ is restricted to the potential region (technically, as an expansion in powers of $q^0/|\mathbf{q}|$), where the internal graviton momentum has the scaling $q\sim(v/r,1/r)$, while the radiated graviton momentum scaling is   $k\sim(v/r,v/r)=\bar{\omega}(1,\boldsymbol{N})$ (with $\bar{\omega}\sim \omega$). Integration in the potential region ensures that from the retarded propagators,
\begin{equation}
    \frac{1}{(q_0+i0)^2-\boldsymbol{q}^2}\rightarrow \frac{1}{v^2(q_0+i0)^2-\boldsymbol{q}^2} \approx -\frac{1}{\boldsymbol{q}^2}+\mathcal{O}(v^2),
\end{equation}
entering in the scattering amplitude, retardation effects only  become important  at order  $\mathcal{O}(v^2)$, which we do not consider here. At subleading  orders in velocities, the amplitude  $\langle M_{5}^{ij}(q,k)\rangle$ has no explicit dependence on $q^{0}$. This takes care of the $q^0$-integration in \eqref{eq:source_some2}, which results in the delta function $\delta(t_{2}-t_{1})$; this can be used to trivialize one of the time integrals\footnote{As a connection with the classical computation, the source multipole moments [given in \eqref{eq:multipolemomentv0}] contain the finite size and retardation effects of the binary, though, at leading and sub-leading orders in velocities, these effects vanish (see e.g., \cite{Blanchet:2013haa}), which is equivalent to the replacement $\tau_{1,2}\rightarrow t$ above.}. With all these simplifications in hand,   the source \eqref{eq:source_some2} becomes
\begin{equation}\label{source}
T^{(0)\,ij}(k,\boldsymbol{z}_1,\boldsymbol{z}_2){=}\frac{i}{m_1 m_2}\int dt \hat{d}^3\boldsymbol{q}e^{i\bar{\omega}\,t{-}i\boldsymbol{q}{\cdot}\boldsymbol{z}_{21}}\langle M_{5,{\rm S}^0}^{(0)\,ij}(\boldsymbol{q},\bar{\omega}){+}M_{5,{\rm S}^1}^{(0)\,ij}(\boldsymbol{q},\bar{\omega}){+}M_{5,{\rm S}^2}^{(0)\,ij}(\boldsymbol{q},\bar{\omega})\rangle {+}\cdots,
\end{equation}
where the amplitude was written in a  spin-multipole decomposition. The superscript $^{(0)}$ indicates that we restrict these to the leading-in-$v$ contribution to the scattering amplitude (See \cref{sec:subleading-v} for the computation at the first sub-leading order in velocities contribution, for spinless BHs). 

\subsection{Scattering amplitude and  double copy}\label{sec:amplitude_double_copy}

With the general formalism to compute the radiated field for the BBH system outlined above, we are left to provide the classical limit of the 5-pt amplitude to be used in the  waveform formula \eqref{eq:radiated field}. The amplitude takes the general form \cite{Bautista:2019tdr}

\begin{equation}\label{eq:newM5clas}
\langle M_{5}^{h}\rangle{=}\frac{- i}{(q{\cdot}k)^{h-1}}\left[\frac{n_h^{(a)}}{(q^{2}{-}q{\cdot}k)(p_{1}{\cdot}k)^{2}}+\frac{n_h^{(b)}}{(q^{2}{+}q{\cdot}k)(p_{2}{\cdot}k)^{2}}\right]\,,\,\,\,
\end{equation}
where $n_h^{(i)}$ are  kinematic  numerators given below. Here we have also included the photon case $h=1$, in order to make contact with existing literature, as well as providing new results at $\mathcal{O}(S^2)$. This formula follows from  the factorization of the 5-pt  amplitude (Figure \ref{fig:5pt-amplitude}) into the product of  a Compton-   and a 3-pt amplitude\footnote{
In \cite{Bautista:2019evw} it was also shown that for gravity, this classical formula follows directly  from  the classical limit of the standard BCJ double copy \cite{Bern:2008qj}, where superclassical contribution to the amplitude can be gauged away by the use of  a \textit{Generalized Gauge Transformation} of the BCJ numerators.}. Then, each kinematic  numerator is computed from the  residue  of the 5-pt amplitude, at the corresponding  pole. Schematically, we have

\begin{figure}[h!]
\begin{equation}\label{cuts m5}
  \includegraphics[width=0.58\textwidth]{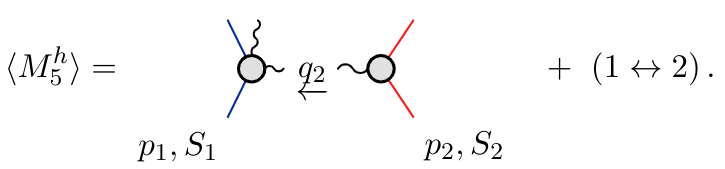}
\end{equation}
\end{figure}
From this factorization, it is clear that the  spurious pole    $q{\cdot} k$ appearing in \eqref{eq:newM5clas} for the gravitational case, $h=2$, arises   from the $t$-channel of the gravitational Compton amplitude, and cancels out in the final result. It was also shown extensively in  \cite{Bautista:2019tdr} that the gravitational numerators in \eqref{eq:newM5clas} can be computed  from a double copy of two photon numerators. That double copy is nothing but the  result of the KLT double copy of 3- and 4-pt amplitudes entering in \eqref{cuts m5},  which can be written symbolically  as $A_n^{(s,\rm{gr})} = K_n A_n^{(0,\rm{em})}\times A_n^{(s,\rm{em})}$ for $n=3,4$; i.e.,  they factorize into the product of the scalar and the spin-$s$ electromagnetic  amplitudes \cite{Bern:2020buy}. Here we make use of this factorization to write out the explicit form of the numerators for both, photon and graviton radiation. 

\subsubsection{Electromagnetic 5-pt amplitude } \label{sec:eletrogmagampl}

We now provide the explicit example for the 5-pt amplitude \eqref{eq:newM5clas}, in the  the electromagnetic case. We start by introducing the  notation for the contractions $F{\cdot}J_{s,A}=F_{\mu\nu}J_{s,A}^{\nu\mu}$, where $F_{\mu\nu}$ is the electromagnetic field strength tensor, and $J_{s,A}^{\nu\mu}$ are the spin-$s$ generators for particle $A$, in Maxwell theory.  Furthermore, we also define the variables 
\begin{align}\label{eq:notation}
 R_{i}^{\mu\nu} =p_{i}^{[\mu}(2\eta_{i}q-k)^{\nu]}, & & \hat{R}_{i}^{\mu\nu }	=2(2\eta_{i}q-k)^{[\mu}J_{s,i}^{\nu]\alpha}(2\eta_{i}q-k)_{\alpha},
\end{align}
with $\eta_{1}=-1$ and $\eta_{2}=1$. With this notation in hand, the scalar numerators for photon radiation read explicitly 
\begin{equation}
    n_{0,\rm{ph}}^{(a)}{=}4e^3 p_1{\cdot}R_2{\cdot}F{\cdot}p_1, \quad 
    n_{0,\rm{ph}}^{(b)}{=}4e^3 p_2{\cdot}R_1{\cdot}F{\cdot}p_2, \label{nphsc}
\end{equation}
where $e$ is the electron charge. This is nothing but the gluing of the numerator of the  spinless electromagnetic Compton amplitude, $A_4\sim p_i{\cdot} F_i{\cdot}F_j{\cdot} p_i$\footnote{ Here $p_i$ is the incoming passive momentum, and  $F_i$ and $F_j$ are the field strength tensors for the incoming and outgoing photons, respectively.},  \cite{Bjerrum-Bohr:2013bxa}, and the scalar 3-pt amplitude $A_3\sim p_i^\mu$, through the photon propagator $\eta_{\mu\nu}$. The triple product notation here corresponds to  the ordered contraction of the Lorentz indices of all components,   $p_i{\cdot} F_i{\cdot}F_j{\cdot} p_i = p_{i,\mu} F_i^{\mu\nu}F_{j,\nu}{}^{\alpha} p_{i,\alpha}$. Furthermore, the  linear-in-spin numerators are
\begin{equation}
\label{eq:lineas-in-spin-numerators-photon}
\begin{split}
  n_{\frac{1}{2},\rm{ph}}^{(a)}	&=n_{0,\rm{ph}}^{(a)}{-}2e^{3}\left[p_{1}{\cdot}R_{2}{\cdot}kF{\cdot}J_{s,1}{-}F_{1q}R_{2}{\cdot}J_{s,1}{+}p_{1}{\cdot}k\,[F,R_{2}]{\cdot}J_{s,1}-p_{1}{\cdot}F{\cdot}\hat{R}_{2}{\cdot}p_{1}\right],\\
n_{\frac{1}{2},\rm{ph}}^{(b)}	&=n_{0,\rm{ph}}^{(b)}{-}2e^{3}\left[p_{2}{\cdot}R_{1}{\cdot}kF{\cdot}J_{s,2}{-}F_{2q}R_{1}{\cdot}J_{s,2}{+}p_{2}{\cdot}k\,[F,R_{1}]{\cdot}J_{s,2}-p_{2}{\cdot}F{\cdot}\hat{R}_{1}{\cdot}p_{2}\right],
\end{split}
\end{equation}
where we introduced the commutator notation $[F,R_{2}]{\cdot}J_{s,i} = (F^\mu{}_{\nu} R_{2}^{\nu\alpha}- R^\mu_{2,\nu} F^{\nu\alpha})(J_{s,i})_{\mu\alpha}  $
These numerators follow analogously  from the  gluing of the electromagnetic spin-$1/2$, 3-pt and 4-pt  amplitudes  \cite{Bautista:2019tdr}. Using  variables \eqref{eq:notation} to rewrite the numerators, trivializes the check for gauge invariance. Notice, on the other hand, the spin contribution in the  $\hat{R}_i$ terms emerges purely  from the linear-in-spin piece of the  3-pt amplitude, whereas the linear-in-spin  Compton amplitude is responsible for   the remaining terms. One can  easily check that by replacing  the numerators \eqref{nphsc} and \eqref{eq:lineas-in-spin-numerators-photon} in the general formula \eqref{eq:newM5clas}, we recover the classical photon  radiation amplitude for the scattering of two colorless charges, with and without spin (compare \cite{Li:2018qap} and \cite{Goldberger:2016iau}, respectively). These numerators have the support of $\delta(p_i{\cdot}(\eta_i q-k))$, which imposes the on-shell condition for the outgoing massive particles in the classical limit\footnote{The classical limit is taken using the $\hbar$-rescaling of the massless momenta, and then taking the leading order for $\hbar\to$.   We have to notice also that in numerators \eqref{eq:lineas-in-spin-numerators-photon} and \eqref{eq:linear-spin-amplitude-gr} below, we have  removed the indices $(\{\alpha\},{ \{\beta\}})$  $(J_{s,i}^{\mu\nu})_{\{\alpha\}}^{\{\beta\}}$ that contract with the  polarization vectors/tensors for the massive particles. In the classical setup, we interpret $J_{s,i}^{\mu\nu}$ simply as the classical spin tensor $S^{\mu\nu}$, which satisfies the spin supplementary condition $p_{\mu}S^{\mu\nu}=0$.}. Finally, the quadratic order in spin numerators are included in the  ancillary \texttt{Mathematica} notebook for this paper. These, again, follow from the electromagnetic quadratic-in-spin 3-pt and 4-pt amplitude, with the latter also included in the ancillary file. For simplicity, at this order we have  restricted to the case in which only one particle has spin, while the other is scalar.

\subsubsection{Gravitational 5-pt amplitude }

Let us now provide the relevant amplitudes to be used in \eqref{eq:radiated field} -- the gravitational amplitudes. We start with the scalar case, for which the gravitational numerators are 
\begin{equation}
\label{eq:scalar-numerators-gr}
    \begin{split}
        n^{(a)}_{0,\rm{gr}}&=\frac{\kappa^3}{4}\left[\left(p_{1}{\cdot}p_{2}F_{1q}-p_{1}{\cdot}k F_{p}\right)^{2}- \frac{m_{1}^{2}m_{2}^{2}}{2}F_{1q}^{2} \right],\\
         n^{(b)}_{0,\rm{gr}}&= -\frac{\kappa^3}{4}\left[\left(p_{1}{\cdot}p_{2}F_{2q}+p_{2}{\cdot}k F_{p}\right)^{2}-\frac{m_{1}^{2}m_{2}^{2}}{2}F_{2q}^{2}\right],
    \end{split}
\end{equation}
where we have used  \textcolor{black}{ $F_{iq}=\eta_{i}(p_{i}{\cdot}F{\cdot}q)$, and $F_p= p_1 {\cdot} F{\cdot} p_2$}. Analogous to the electromagnetic numerators, this can also be obtained from the gluing of the scalar, 3-pt and 4-pt amplitudes through the graviton propagator.    These numerators  can  be introduced in the general formula \eqref{eq:newM5clas}, to  recover the result for the classical limit of the gravitational  amplitude for scalar particles \cite{Goldberger:2016iau,Luna:2017dtq,Bautista:2019evw}. Next, the gravitational numerators to linear-order  in spin  can analogously be computed to get 
\begin{equation}\label{eq:linear-spin-amplitude-gr}
\begin{split}
    n^{(a)}_{\frac{1}{2},\rm{gr}}& = \frac{\kappa^3}{8}\Big\{\left(p_{1}{\cdot}p_{2}F_{1q}-p_{1}{\cdot}k F_{p}\right) \left[\left(  p_1{\cdot} p_2\,q{\cdot} k+p_1{\cdot} k\,p_2{\cdot} k   \right)   F{\cdot}J_{2s,1}{-}F_{1q}R_{2}{\cdot}J_{2s,1}{+}p_{1}{\cdot}k\,[F,R_{2}]{\cdot}J_{2s,1}\right]\\
     &\hspace{1.5cm}+\frac{m_{2}^{2}F_{1q}}{2}\left[F_{1q}(2q{-}k){\cdot}J_{2s,1}{\cdot}p_{1}{-}m_{1}^{2}q{\cdot}k\,F{\cdot}J_{2s,1}+p_{1}{\cdot}k(2q{-}k){\cdot}F{\cdot}J_{2s,1}{\cdot}p_{1}\right]\Big\},\\
      n^{(b)}_{\frac{1}{2},\rm{gr}} & =- \frac{\kappa^3}{8}\Big\{ \left(p_{1}{\cdot}p_{2}F_{2q}+p_{2}{\cdot}k F_{p}\right)
       \left(F_{2q} (2q+k){\cdot}J_{2s,1}{\cdot}p_2 -  p_2{\cdot}k\,p_2{\cdot}F{\cdot}J_{2s,1}{\cdot}(2q+k)\right) \\
     &\hspace{1.5cm} +\frac{m_{2}^{2}F_{2q}^{2}}{2}(2q{+}k){\cdot}J_{2s,1}{\cdot}p_{1}\Big\}.
    \end{split} 
\end{equation}
Here we point out that the generators $J_{2s,1}$ act in the gravity theory rather than their electromagnetic counterpart.  Similarly to the scalar case, these numerators can be placed in \eqref{eq:newM5clas} to recover the corresponding gravitational amplitude. To obtain the full amplitude for both particles with spin, we utilize the symmetrization mappings 
\begin{align}
m_1\leftrightarrow m_2, & & p_1\leftrightarrow p_2, & & q\rightarrow-q, & & J_{2s,1}\rightarrow J_{2s,2},
\end{align}
in the final formula. The resulting amplitude  recovers the spinning amplitude in dilaton gravity computed in \cite{Li:2018qap} for classical spinning sources, once the  terms proportional to $m_i$ in the  numerators in \eqref{eq:linear-spin-amplitude-gr}, which arise  from the graviton propagator, are removed. 

Analogous to the electromagnetic case in \cref{sec:eletrogmagampl}, at the quadratic order in spin, for simplicity, we restrict ourselves to the scenario, where only one BH is spinning, $S_2\rightarrow 0$. We include the numerators as well as the gravitational Compton amplitude in the ancillary \texttt{Mathematica} notebook. We have checked that up to a contact term, which is irrelevant for the gravitational waveform, the quadratic-in-spin amplitude recovers the results in \cite{Jakobsen:2021lvp}\footnote{We would like to thank Gustav Mogull et al. for sharing their results with us before publication.}.

\subsection{Computation of the radiated field}\label{sec:radiated_field_amplitudes}

In the previous sections, we built up the 5-pt gravitational spinning scattering amplitude up to quadratic order in the BHs' spins. With this, we can now return to \eqref{eq:radiated field} to successively construct the emitted classical gravitational radiation from the spinning BBH at increasing PN order. First, we compute the gravitational waveform to leading order in velocity up to quadratic order in the BHs' spins, while turning to the  computation of the waveform  at sub-leading order in the BHs' velocities in the spin-less limit in  \cref{sec:subleading-v}.

\subsubsection{Scalar waveform}\label{sec:scalar_waveform}

The derivation of the Einstein quadrupole formula from scattering amplitudes was first done by Hari Dass and Soni in \cite{HariDass:1980tq}; more  recently, it was derived by Goldberger's and Rigway's classical double copy approach \cite{Goldberger:2017vcg}. In the following, we re-derive the scalar term of the  waveform in the Goldberger and Rigway setup, for completeness. This in turn, will outline the formalism used throughout the remaining sections to arrive at the corrections to the quadrupole formula.  Expanding the scalar amplitude \eqref{eq:scalar-numerators-gr} to leading order in velocities $v$, we find  
\begin{equation}\label{eq:scalar-amplitude}
\langle M_{5,{\rm {S}^0}}^{(0)\,ab}(\boldsymbol{q},\bar{\omega})\rangle=-i\frac{m_{1}^{2}m_{2}^{2}}{4}\kappa^{3}\left[2\frac{q^{a}q^{b}}{\boldsymbol{q}^4}+\frac{1}{\bar{\omega}\boldsymbol{q}^2}\left(q^{a}v_{12}^{b}+q^{b}v_{12}^{a}\right)\right],
\end{equation}
where $v_{AB}=v_A-v_B$. Substituting this amplitude into the  scalar source \eqref{source}, and integrating over $\boldsymbol{q}$ using \eqref{eq:integrals}, the non-spinning source reduces to
\begin{equation}\label{eq:scalar-source-befor-eom}
T_{\rm{S}^0}^{(0)\,ab}(k,\boldsymbol{z}_1,\boldsymbol{z}_2){=}{ -}\int dte^{i\bar{\omega}\,t}\frac{\kappa^{3}}{32\pi}\sum_{A,B}\frac{m_{A}m_{B}}{r^{3}}\left[\left(z_{AB}^{a}z_{AB}^{a}  {-}r^2\delta^{ab} \right)+\frac{2i}{\bar{\omega}}\left(z_{AB}^{a}v_{A}^{b}{+}z_{AB}^{a}v_{A}^{a}\right)\right].
\end{equation} 
Here, and in the following, single label sums are understood to run over the two massive particle labels, $\sum_A:=\sum_{A=1}^2$, while the double sum  is performed imposing the constraint $A\neq B$: $\sum_{A,B}:=\sum_{A\neq B;A,B=1}^2$.

Notice that the term proportional to $\delta^{ab}$ in \eqref{eq:scalar-source-befor-eom} vanishes under the action of the TT-projector in \eqref{eq:radiated field}. Therefore, in the following, we remove this term from the source and focus only on those parts contributing non-trivially to the TT radiated field. Now, we use the non-spinning part of the EoM \eqref{eq:EoM} to rewrite the second term in the square bracket of \eqref{eq:scalar-source-befor-eom}:
\begin{equation}\label{eq:scalar-source-intermediate}
T_{\rm{S}^0}^{(0)\,ab}(k,\boldsymbol{z}_1,\boldsymbol{z}_2)=-\kappa\int dte^{i\bar{\omega}\,t} \left[\sum_{A,B}\frac{\kappa^{2}m_{A}m_{B}}{32\pi}\frac{z_{AB}^{a}z_{AB}^{a}}{r^{3}}-\frac{2i}{\bar{\omega}}\sum_{A}m_{A}\left(v_{A}^{b}\dot{v}_{A}^{a}+v_{A}^{a}\dot{v}_{A}^{b}\right)\right].
\end{equation}
The second term of this expression can be further integrated, since $v_{A}^{b}\dot{v}_{A}^{a}+v_{A}^{a}\dot{v}_{A}^{b}=\frac{d}{dt}\left(v_{A}^{a}v_{A}^{b}\right)$. As for the first term, this can be rewritten using
\begin{equation}\label{eq:identity_eom}
\frac{\kappa^{2}}{32\pi}\sum_{A,B}m_{A}m_{B}\frac{z_{AB}^{a}z_{AB}^{a}}{r^{3}}=-\sum_{A}m_{A}\left(\ddot{z}_{A}^{a}z_{A}^{b}+z_{A}^{a}\ddot{z}_{A}^{b}\right),
\end{equation}
derived from the scalar EoM. Putting these ingredients together into \eqref{eq:scalar-source-intermediate}, we find the scalar source to be
\begin{equation}
T_{\rm{S}^0}^{(0)\,ab}(k,\boldsymbol{z}_1,\boldsymbol{z}_2)=\kappa\int dte^{i\bar{\omega}\,t}\sum_{A} m_{A}\left(\ddot{z}_{A}^{a}z_{A}^{b}+z_{A}^{a}\ddot{z}_{A}^{b}+2v_{A}^{a}v_{A}^{b}\right).
\end{equation}
Using the relation $2v_{A}^{a}v_{A}^{b}=\frac{d^{2}}{dt^{2}}\left(z_{A}^{a}z_{A}^{b}\right)-\left(\ddot{z}_{A}^{a}z_{A}^{b}+z_{A}^{a}\ddot{z}_{A}^{b}\right)$, the above expression can be put into the more compact form
\begin{equation}
T_{\rm{S}^0}^{(0)\,ab}(k,\boldsymbol{z}_1,\boldsymbol{z}_2)= \kappa \int dte^{i\bar{\omega}\,t}\sum_{A}m_{A}\frac{d^{2}}{dt^{2}}\left(z_{A}^{a}z_{A}^{b}\right),
\label{eq:scalarsource}
\end{equation}
which in turn implies that the radiated field \eqref{eq:radiated field} for a non-spinning BBH takes the familiar Einstein quadrupolar form:
\begin{equation}\label{eq:einstein-quadrupole}
\boxed{
h_{TT,\,\rm{S}^0}^{(0)\,ij}(T_{R},R,\boldsymbol{N} ,\boldsymbol{z}_1,\boldsymbol{z}_2)= \frac{\kappa^2}{16\pi R}\Pi^{ij}{}_{ab}\sum_{A} m_{A}\left[\frac{d^{2}}{dt^{2}}\left(z_{A}^{a}z_{A}^{b}\right)\right]_{t=T_{R}}.}
\end{equation}
The sequence of Fourier transforms in the source \eqref{eq:scalarsource} and \eqref{eq:radiated field} leads to the evaluation of the emitted gravitational radiation at retarded time $T_R$, therefore, recovering the classical result \eqref{eq:classicalhijgeneralorbit} in the no-spin-limit. As a quick remark,  notice when restoring Newton's constant $G$ the quadrupole radiation is linear in $G$, as opposed to gravitational Bremsstrahlung, which is  quadratic \cite{Peters:1970mx,1985Konradin,1977KT,1978KT}. This is of course just a  feature of using the EoM to rewrite the source. 

\subsubsection{Linear-in-spin waveform}\label{sec:linear_spin_waveform}

In the previous section, the main components of the derivation of the gravitational waveform from a compact binary system were outlined. In particular, we have seen that the classical EoM play an important role in recovering the quadrupole formula. Going beyond this, at linear order in the BHs' spins, there are two contributions to the waveform. First, the scalar amplitude could be iterated with the linear-in-spin part of the classical EoM \eqref{eq:EoM}; this contribution, however, is sub-leading in velocity as made explicit in \eqref{eq:EoM}. Secondly, the linear-in-spin amplitude, in conjunction with the non-spinning part of the EoM gives rise to a leading in BHs' velocities and linear-in-their spins contribution to the waveform. To determine the latter, we start from the linear-in-spin amplitude \eqref{eq:linear-spin-amplitude-gr}, where the leading in $v$ expression is given by
\begin{equation}\label{eq:linear_spin_v0}
   \langle M_{5,{\rm S}^{1}}^{(0)\,ab}(\boldsymbol{q},\bar{\omega})\rangle=-\frac{m_{1}m_{2}\kappa^{3}}{8}\varepsilon_{efk}\left(m_{2}S_{1}^{k}{-}m_{1}S_{2}^{k}\right)N^{[e}\left(\delta^{f]a}\delta^{bc}{+}\delta^{f]b}\delta^{ac}\right)\frac{q^{c}}{\boldsymbol{q}^{2}}.
\end{equation}
Analogous to the scalar case, we can substitute this amplitude into  \eqref{source} to get the  linear-in-spin source $T_{{\rm {S}^{1}}}^{(0)\,ab}$. After integrating over $\bs{q}$, utilizing \eqref{eq:integrals}, this source simplifies to
\begin{equation}
    T_{{\rm {S}^{1}}}^{(0)\,ab}(k,\boldsymbol{z}_{1},\boldsymbol{z}_{2})=\frac{\kappa^{3}}{32\pi}\varepsilon_{efk}\left(m_{2}S_{1}^{k}{-}m_{1}S_{2}^{k}\right)N^{[e}\left(\delta^{f]a}\delta^{bc}{+}\delta^{f]b}\delta^{ac}\right)\int dte^{i\bar{\omega}\,t}\frac{z_{21}^{c}}{r^{3}}.
\end{equation}
Powers of $r$ in the denominator can be removed by using the scalar limit of the classical EoM \eqref{eq:EoM}. Then, analogous to the scalar computation, the linear-in-spin source is
\begin{equation}
T_{{\rm {S}^{1}}}^{(0)\,ab}(k,\boldsymbol{z}_{1},\boldsymbol{z}_{2})=\kappa\varepsilon_{efk}S_{1}^{k}N^{[e}\left(\delta^{f]a}\delta^{bc}{+}\delta^{f]b}\delta^{ac}\right)\int dte^{i\bar{\omega}\,t}\dot{v}_{1}^{c}+(1\leftrightarrow2).
\label{eq:spin1source}
\end{equation}
Finally, the linear in spin corrections to the Einstein quadrupole formula, derived from the above amplitude, obtained from \eqref{eq:spin1source}, together with \eqref{eq:radiated field}, are
\begin{equation}\boxed{h_{TT,\,{\rm {S}^{1}}}^{(0)\,ij}(T_{R},R,\boldsymbol{N},\boldsymbol{z}_1,\boldsymbol{z}_2)=\frac{\kappa^{2}}{16\pi R}\Pi^{ij}{}_{ab}\varepsilon_{efk}\sum_{A}S_{A}^{k}\left[N^{[e}\left(\delta^{f]a}\delta^{bc}{+}\delta^{f]b}\delta^{ac}\right)\dot{v}_{A}^{c}\right]\Big|_{T_{R}}.}
\label{eq:amphijS1}
\end{equation}
At this stage, this correction is valid, similar to the quadrupole formula, for general closed orbits. We find a perfect match of these spinning corrections at linear order in the objects' spins, with the classical derivation, \eqref{eq:classicalhijgeneralorbit}, using the identity \eqref{eq:identityTT}. The linear-in-spin scattering amplitude is universal \cite{Bjerrum-Bohr:2013bxa,Bautista:2019tdr}, therefore, so is the radiated gravitational field \eqref{eq:amphijS1}. Equivalently, the classical spin dipole of a point particle is universal, describing any spinning compact object at leading order. Therefore, non-universality of the waveform at higher spin orders may enter only through a solution to the classical EoM for a particular compact binary system. We showed in \cref{sec:hijclassical} that the closed orbits waveform \eqref{eq:classicalhijgeneralorbit} contains all possible spin effects at leading order in the BHs' velocities, \textit{before} specializing the constituents' trajectories; i.e., $h^{(0),ij}_{TT,S^{\ell\geq 2}}=0$. Therefore, we expect to find cancellations at higher orders in spins at the level of the scattering amplitude for $\ell> 1$. Finally, as claimed above, there exists a one-to-one correspondence between source multipole moments and spinning scattering amplitudes: $\mc{I}_{ij}\leftrightarrow \langle M_{5,{\rm S}^{0}}^{(0)\,ab}\rangle$ and $\mc{J}_{ij}\leftrightarrow \langle M_{5,{\rm S}^{1}}^{(0)\,ab}\rangle$. This holds in the sense that both $\mc{I}_{ij}$ and $\langle M_{5,{\rm S}^{0}}^{(0)\,ab}\rangle$ produce the quadrupole formula (and similarly for the linear-in-spin waveform).

\subsubsection{Cancellations at quadratic order in spin}\label{sec:quadratic in spin}

In the previous section, we showed that the gravitational waveform emitted from a spinning BBH at leading order in its velocities is entirely contained in the linear-in-spin radiation field \eqref{eq:classicalhijgeneralorbit}. Equivalently, this waveform is obtained only using the scalar and linear-in-spin amplitude. The remaining all orders in spin result \eqref{eq:hijcircularv0} emerges solely from the solution \eqref{eq:consvsol} for quasi-circular orbits. To confirm this from the scattering amplitudes perspective, we are left to show that higher spin amplitudes do not provide additional non-trivial contributions to the general closed orbit results presented above. In this section, we demonstrate the cancellation at the quadratic order in the BHs' spins, by specializing to circular orbits and by focusing on the $S_1\neq 0, S_2\rightarrow0$ limit.

At leading order in the BHs' velocities, there are two distinct contributions to the radiated field from our approach. There is the quadratic-in-spin part of the amplitude on the one hand (which we provide in the ancillary \texttt{Mathematica} notebook for brevity), leading to $T_{1,{\rm {S}^{2}}}^{(0)\,ij}$, and the scalar part \eqref{eq:scalar-amplitude} in conjunction with the quadratic-in-spin part of the classical EoM \eqref{eq:EoM}, yielding $T_{2,{\rm {S}^{2}}}^{(0)\,ij}$, on the other hand\footnote{Notice, the linear-in-spin part of the EoM is sub-leading in $v$, and therefore, when convoluted with the linear-in-spin amplitude, the resulting quadratic in spin contribution is pushed to sub-leading order in velocities.}; both combine as
\begin{align}
    T_{\rm {S}^{2}}^{(0)\,ij}(k,\boldsymbol{z}_{1},\boldsymbol{z}_{2})=T_{1,{\rm {S}^{2}}}^{(0)\,ij}(k,\boldsymbol{z}_{1},\boldsymbol{z}_{2})+T_{2,{\rm {S}^{2}}}^{(0)\,ij}(k,\boldsymbol{z}_{1},\boldsymbol{z}_{2}).
\end{align}
Focusing first on the contribution from the quadratic-in-spin part of the  amplitude, to leading order in $v$ it reads 
\begin{equation}\label{eq:amplitude spin 2 nonrel}
    \langle M_{5,{\rm S}^{2}}^{(0)\,ab}(\boldsymbol{q},\bar{\omega})\rangle= \textcolor{black}{\frac{1}{4}\,}i m_{2}^{2}\kappa^{3}S_{1}^{k}S_{1}^{l}\left[V_{kl,df}^{ab}\frac{q^{d}q^{f}}{\boldsymbol{q}^{2}}+C_{kl}^{ab}\right],
\end{equation}
where we have defined the tensor 
$ V_{kl,df}^{ab} = \delta_{kl}\delta_{d}^{a}\delta_{f}^{b}-\frac{1}{2}\delta_{kd}\left(\delta_{f}^{a}\delta_{l}^{b}+\delta^{fb}\delta_{l}^{a}\right)$, and $C_{kl}^{ab}$ is a contact term, which we discard, as it is irrelevant for the gravitational waveform. As before, we insert this amplitude into the source \eqref{source}, and perform the $\bs{q}$-integrals aided by \eqref{eq:integrals}. The first contribution to the source $T_{\rm {S}^{2}}^{(0)\,ij}$ is then
\begin{equation}
    T_{1,{\rm {S}^{2}}}^{(0)\,ab}(k,\boldsymbol{z}_{1},\boldsymbol{z}_{2})=-\textcolor{black}{\frac{1}{4}\,}\frac{m_{2}\kappa^{3}}{m_14\pi}S_{1}^{k}S_{1}^{l}V_{kl,df}^{ab}\int dte^{i\bar{\omega}\,t}\frac{1}{r^{5}}\left[r^{2}\delta^{df}-3z_{21}^{d}z_{21}^{f}\right].
\end{equation}
Using the scalar part of the  EoM \eqref{eq:EoM} to remove three powers of $r$ in the denominator, the above reduces to
\begin{equation}\label{eq:quad source quad ampl}
    T_{{1,\rm {S}^{2}}}^{(0)\,ab}(k,\boldsymbol{z}_{1},\boldsymbol{z}_{2})=-3\frac{m_{2}}{m_1}\text{\ensuremath{\kappa}}S_{1}^{k}S_{1}^{l}V_{kl,df}^{ab}\int dte^{i\bar{\omega}\,t}\frac{1}{r^{2}}\left[\left(\frac{\dot{v}_{2}{\cdot}z_{12}}{m_{1}}{+}\frac{\dot{v}_{1}{\cdot}z_{21}}{m_{2}}\right)\frac{\delta^{df}}{3}{-}\left(\frac{\dot{v}_{2}^{(d}z_{12}^{f)}}{m_{1}}{+}\frac{\dot{v}_{1}^{(d}z_{21}^{f)}}{m_{2}}\right)\right], 
\end{equation}
which, for quasi-circular orbits \eqref{eq:circorbits}, reads
\begin{equation}\label{eq:cirspin2amp2}
     T_{1,{\rm {S}^{2}}}^{(0)\,ab}(k,\boldsymbol{z}_{1},\boldsymbol{z}_{2})\Big|_{\rm{circular}} = \textcolor{black}{-2}\kappa\bar{\omega}^2\mu \bar{a}_1^2\int dte^{i\bar{\omega}t}\left[2n^a n^b - \lambda^a \lambda^b\right].
\end{equation}
Recall the definition for the symmetric mass ratio $\mu = m_1 m_2/M$, and  $\bar{a}_1 = S_1^i \ell_i/m_1$, with $\ell^i$  perpendicular to both $n^i$ and $\lambda^i$. Note, the solution to the classical EoM, $r(x)$, in the numerator, cancels with the two powers of $r$ in the denominator.

We now turn to the second contribution to the source: $T_{2,\rm {S}^{2}}^{(0)\,ij}$. To that end, we first rewrite \eqref{eq:scalar-source-befor-eom} by expanding the sums and removing those terms that vanish under the TT projection:
\begin{equation}
    T_{{2,\rm {S}^{2}}}^{(0)\,ab}(k,\boldsymbol{z}_{1},\boldsymbol{z}_{2})= -\kappa^3 \int dt e^{i\bar \omega t}\frac{m_1 m_2 z_{12}^c}{16\pi r^3}\left[  \delta ^{c(a} z_{12}^{b)}+\frac{2i}{\bar \omega }\delta^{c(a}v_{12}^{b)}\right].
\label{eq:t2quadratic}
\end{equation}
Next we use the classical EoM to quadratic order in spin, which can be written in the following form (see Appendix \ref{sec: eom quad spin})
\begin{equation}\label{eq:eom_wuad_spin}
     \dot{v}_1^l = {-}\frac{m_2\kappa^2}{32\pi}  \frac{z_{12}^l}{r^3}{\textcolor{black}{+}}\frac{3}{4}\frac{S_1^iS_1^j}{m_1^2r^2}\left[\left(\delta_{ij}{-}\frac{5z_{12,i}z_{12,j}}{r^2}\right)\left(\dot{v}_1^l {-}\frac{m_2}{m_1}\dot{v}_2^l  \right){+}2 \delta^l_{(i}\left(\dot{v}_{1,j)}{-}\frac{m_2}{m_1} \dot{v}_{2,j)}  \right) \right].
\end{equation}
Combining this with \eqref{eq:t2quadratic}, the scalar part will recover the Einstein quadrupole radiation formula \eqref{eq:einstein-quadrupole}. We stress that although the quadrupole formula appears to be spin-independent for general orbits, spin information arises through a specific solution to the EoM, as pointed out above. In particular, for quasi-circular orbits the Einstein quadrupole formula provides the quadratic-in-spin result \eqref{eq:hijcircularv0}. Let us, therefore, focus in the remaining contribution of \eqref{eq:eom_wuad_spin}, which is 
\begin{equation}
\begin{split}
T_{{2,{\rm {S}^{2}}}}^{(0)\,ab}(k,\boldsymbol{z}_{1},\boldsymbol{z}_{2}) & =-\frac{\textcolor{black}{3}}{4}\kappa S_{1}^{k}S_{1}^{l}\int dte^{i\bar{\omega}t}\left[\delta^{c(a}z_{12}^{b)}+\frac{2i}{\bar{\omega}}\delta^{c(a}v_{12}^{b)}\right]\times\frac{1}{m_{1}r^{2}}\\
 &  \left[\left(\delta_{kl}{-}\frac{5z_{12,k}z_{12,l}}{r^{2}}\right)\left(\dot{v}_{1}^{c}{-}\frac{m_{2}}{m_{1}}\dot{v}_{2}^{c}\right){+}2\delta_{(k}^{c}\left(\dot{v}_{1,l)}{-}\frac{m_{2}}{m_{1}}\dot{v}_{2,l)}\right){+}\frac{m_{2}}{m_{1}}(1\leftrightarrow2)\right].
\end{split}
\end{equation}
Using the center of mass parametrization\footnote{Note, the linear-in-spin corrections of this parametrization is sub-leading in velocities.} \eqref{COMframe}, the quasi-circular orbits condition $\ddot{r} = -\bar{\omega}r$, and the unit vectors \eqref{eq:circorbits}, the source reduces to
\begin{equation}
    T_{{2,{\rm {S}^{2}}}}^{(0)\,ab}(k,\boldsymbol{z}_{1},\boldsymbol{z}_{2})\Big|_{\rm{circular}} =\textcolor{black}{3}\kappa\bar{\omega}^2 \mu \bar{a}_1^2\int dte^{i\bar{\omega}t}\left[n^a n^b + i (\lambda^a n^b+\lambda^b n^a)\right],
\end{equation}
In order to remove the imaginary part of the source, we proceed as before and use an IBP prescription. Notice, since $(\lambda^a n^b+\lambda^b n^a) = -\frac{1}{\omega}\frac{d}{dt}(\lambda^a\lambda^b)$, the IBP yields
\begin{equation}\label{eq:source quad spin scalar ampl}
    T_{{2,{\rm {S}^{2}}}}^{(0)\,ab}(k,\boldsymbol{z}_{1},\boldsymbol{z}_{2})\Big|_{\rm{circular}} =\textcolor{black}{3}\kappa\bar{\omega}^2\mu \bar{a}_1^2\int dte^{i\bar{\omega}t}\left[n^a n^b - \lambda^a \lambda^b\right].
\end{equation}
This has the familiar form found in \eqref{eq:hijcircularv0}. Unlike this form, in \eqref{eq:cirspin2amp2} an extra factor of two appears in the $n^an^b$ term. This obscures the desired cancellation between \eqref{eq:source quad spin scalar ampl} and \eqref{eq:cirspin2amp2} in $T_{\rm {S}^{2}}^{(0)\,ij}$. To address this subtlety, we emphasize the degeneracy in choice of the IBP prescription. For instance, the relations of the kinematic variables in the center of mass frame results in $-\frac{d}{dt}(\lambda^a\lambda^b) = \frac{d}{dt}(n^a n^b)=\omega(\lambda^a n^b+\lambda^b n^a)$. Using the latter equality, the IBP performed in \eqref{eq:source quad spin scalar ampl} results in $2n^a n^b$, instead of $n^a n^b - \lambda^a \lambda^b$. A priori, neither of these two choices are preferred. The solution is to notice that the freedom in the choice of the IBP prescription is a manifestation of the gauge redundancy of the gravitational waveform at null infinity. That is, below in \cref{sec:results} we show that either choice yields the same result for the gauge invariant gravitational wave energy flux. For now, we note only that at the level of the gauge invariant energy flux, one factor of $n^a n^b $ in  \eqref{eq:cirspin2amp2} is equivalent to $n^a n^b\to \frac{1}{2}(n^a n^b - \lambda^a \lambda^b)$, and postpone the justification to \cref{sec:results}. Therefore, both \eqref{eq:cirspin2amp2} and \eqref{eq:source quad spin scalar ampl} yield the same result, but with opposite sign. This implies the desired cancellation of the waveform contributions at the quadratic order in BHs' spins. Equivalently, using the waveform derived from \eqref{eq:cirspin2amp2} and \eqref{eq:source quad spin scalar ampl} to determine the energy flux from each contribution, we see that both contributions are identical up to an overall sign, hence, cancelling at the level of the gauge invariant gravitational wave energy flux as well (more on this below).

\subsubsection{Scalar waveform at sub-leading order in velocities} \label{sec:subleading-v}

So far we have dealt with leading in BHs' velocities spinning corrections to the Einstein quadrupole formula \eqref{eq:einstein-quadrupole}. In this section, we go beyond this restriction and consider a non-spinning BBH at the first sub-leading order in velocities, therefore, demonstrating the applicability of our approach \eqref{eq:radiated field} to determine the radiated gravitational waves also in this regime. At this order, the scalar 5-pt amplitude is also independent of $q^0$, therefore, arguments made above in \cref{sec:general_approach_amp} concerning the time integration still holds. In this case, however, the first relativistic correction, in our    Born approximation, as coming from the product of the plane wave functions in \eqref{eq:source_some2}, appears in the source through the     kinematic exponential $e^{-i\boldsymbol{ k}\cdot\tilde{\bs{z}}}$, and therefore contributes to the sub-leading source $T_{S^0}^{(1)\,ab}$, due to the scaling $\bar{\omega}\sim v$. That is, after time integration, the exponential function reduces as $e^{i \bs{k}\cdot\tilde{\bs{z}}}\rightarrow 1-i\bar{\omega} \boldsymbol{N}{\cdot} \tilde{\boldsymbol{z}}+\mathcal{O}(v^2)= 1-\frac{i}{2}\bar{\omega} \boldsymbol{N}{\cdot}( \boldsymbol{z}_1+\boldsymbol{z}_2)+\mathcal{O}(v^2)$; hence, the source is built from the order-$v^0$ non-spinning scattering amplitude, $T_{2,S^0}^{(1)\,ab}$, as well as from the $v^1$-amplitude, $T_{1,S^0}^{(1)\,ab}$. More concretely, the sub-leading source decomposes as\footnote{In principle, the classical EoM \eqref{eq:EoM} also contain higher-order-in-$v$ corrections, which could be used in an iterative manner, starting purely from the leading in $v$-scalar amplitude. However, these velocity corrections vanish in the no-spin limit considered in this section.}
\begin{equation}\label{eq:sourcev1gen}
    T_{S^0}^{(1)\,ab}(k,\boldsymbol{z}_1,\boldsymbol{z}_2) = T_{1,S^0}^{(1)\,ab}(k,\boldsymbol{z}_1,\boldsymbol{z}_2)+T_{2,S^0}^{(1)\,ab}(k,\boldsymbol{z}_1,\boldsymbol{z}_2),
\end{equation}
where
\begin{equation}\label{eq:soruce1}
T_{1,S^0}^{(1)\,ab}(k,\boldsymbol{z}_1,\boldsymbol{z}_2)=\frac{i}{m_1 m_2}\int dte^{i\bar{\omega}\,t}\int\frac{d^{3}\boldsymbol{q}}{\left(2\pi\right)^{3}}e^{-i\boldsymbol{q}{\cdot}\boldsymbol{z}_{21}}\langle M_{5,{\rm S}^0}^{(1)\,ab}(\boldsymbol{q},\bar{\omega})\rangle ,
\end{equation}
and
\begin{equation}\label{eq:soruce2}
T_{2,S^0}^{(1)\,ab}(k,\boldsymbol{z}_1,\boldsymbol{z}_2)=\frac{1}{m_1 m_2}\int dte^{i\bar{\omega}\,t}\int\frac{d^{3}\boldsymbol{q}}{\left(2\pi\right)^{3}}e^{-i\boldsymbol{q}{\cdot}\boldsymbol{z}_{21}}\frac{\bar{\omega}}{2} \boldsymbol{N}{\cdot}( \boldsymbol{z}_1+\boldsymbol{z}_2)\langle M_{5,{\rm S}^0}^{(0)\,ab}(\boldsymbol{q},\bar{\omega})\rangle.
\end{equation}
Notice the superscripts in the amplitude. First, we focus on the relativistically corrected scalar amplitude. Analogous to before, we insert \eqref{eq:scalar-numerators-gr} into \eqref{eq:newM5clas}, but now keep the non-trivial order $\mathcal{O}(v^1)$ contributions to the 5-pt amplitude:
\begin{equation}
\begin{split}
\langle M_{5,{\rm S^0}}^{(1)\,ab}(\boldsymbol{q},\bar{\omega})\rangle & =-\frac{i m_{1}^{2}m_{2}^{2}\kappa^{3}}{2}N_{l}\Bigg[\frac{q^{l}q^{m}}{\boldsymbol{q}^{4}}\delta_{m}^{(a}(v_{1}{+}v_{2})^{b)}+\frac{q^{l}}{2\bar{\omega}\boldsymbol{q}^{2}}\left(v_{1}^{a}v_{1}^{b}-v_{2}^{a}v_{2}^{b}\right)\\
 & \hspace{3cm}+\frac{q^{m}}{\bar{\omega}\boldsymbol{q}^{2}}\left(v_{1}^{l}v_{1}^{(a}\delta_{m}^{b)}-v_{2}^{l}v_{2}^{(a}\delta_{m}^{b)}\right)\Bigg].
 \end{split}
\end{equation}
Subsequently, the source \eqref{eq:soruce1}, after the $\bs{q}$-integration, takes the form
\begin{equation}
\begin{split}
    T_{1,S^0}^{(1)\,ab}(k,\boldsymbol{z}_1,\boldsymbol{z}_2) & =\frac{\kappa^{3}}{16\pi}N_{l}\int dte^{i\bar{\omega}\,t}\sum_{A,B}\frac{m_{A}m_{B}}{r^{3}}\Bigg[\frac{1}{2}\left(\boldsymbol{z}_{AB}^{2}\delta^{lm}{-}z_{AB}^{l}z_{AB}^{m}\right)\delta_{m}^{(a}(v_{A}{+}v_{B})^{b)}\\
 & \hspace{3cm}{-}\frac{i}{\bar{\omega}}z_{AB}^{n}\left(\delta_{n}^{l}v_{A}^{a}v_{A}^{b}{+}2\delta_{n}^{m}v_{A}^{l}v_{A}^{(a}\delta_{m}^{b)}\right)\Bigg].
 \end{split}
\end{equation}
In order to remove the powers of $z_{AB}\sim r$ in the denominator, we use the scalar part of the EoM \eqref{eq:EoM}, to obtain
\begin{align}
\begin{split}
T_{1,S^0}^{(1)\,ab}(k,\boldsymbol{z}_1,\boldsymbol{z}_2) & =2\kappa N_{l}\int dte^{i\bar{\omega}\,t}\Bigg[-\frac{1}{2}\sum_{A,B}m_{A}\left(\dot{\boldsymbol{v}}_{A}{\cdot}\boldsymbol{z}_{AB}\delta^{lm}{-}\dot{v}_{A}^{m}z_{AB}^{l}\right)\delta_{m}^{(a}(v_{A}{+}v_{B})^{b)}\\
 & \hspace{3cm}{+}\frac{i}{\bar{\omega}}\sum_{A} m_{A}\dot{v}_{A}^{n}\left(\delta_{n}^{l}v_{A}^{a}v_{A}^{b}{+}2\delta_{n}^{m}v_{A}^{l}v_{A}^{(a}\delta_{m}^{b)}\right)\Bigg].
\end{split}
\end{align}
The term in the second line can be integrated utilizing the relation $\dot{v}_{A}^{n}\left(\delta_{n}^{l}v_{A}^{i}v_{A}^{j}{+}2\delta_{n}^{m}v_{A}^{l}v_{A}^{(i}\delta_{m}^{j)}\right)= \frac{d}{dt}(v_{A}^i v_{A}^j v_{A}^l)$. With this, this piece of the sub-leading scalar source simplifies to
\begin{align}
\begin{split}
  T_{1,S^0 }^{(1)\,ab}(k,\boldsymbol{z}_1,\boldsymbol{z}_2) & =-2\kappa N_{l}\int dte^{i\bar{\omega}\,t}\Bigg[\sum_{A,B}\frac{m_{A}}{2}\left(\dot{\boldsymbol{v}}_{A}{\cdot}\boldsymbol{ z}_{AB}\delta^{lm}{-}\dot{v}_{A}^{m}z_{AB}^{l}\right)\delta_{m}^{(a}(v_{A}{+}v_{B})^{b)}\\
  &\hspace{4cm}
  {-}\sum_{A} m_{A}v_{A}^{l}v_{A}^{a}v_{A}^{b}\Bigg].
\end{split}
\end{align}
We now address the second term in \eqref{eq:sourcev1gen} -- the computation of the second contribution \eqref{eq:soruce2} to the sub-leading scalar source. The $\bs{q}$-integration is identical to the one used leading up to \eqref{eq:scalar-source-intermediate}. Starting from the latter, using the relation \eqref{eq:identity_eom}, and multiplying the sub-leading prefactor $-\frac{i}{2}\bar{\omega}\boldsymbol{N}{\cdot}(\boldsymbol{z}_1+\boldsymbol{z}_2)$ we arrive at
\begin{equation}
T_{2,S^0}^{(1)\,ab}(k,\boldsymbol{z}_1,\boldsymbol{z}_2)
=-\kappa\int dte^{i\bar{\omega}\,t}\boldsymbol{N}{\cdot}(\boldsymbol{z}_1+\boldsymbol{z}_2)\sum_A m_{A}\left[i\bar{\omega}\ddot{z}_{A}^{(a}z_{A}^{b)}-2v_{A}^{(a}\dot{v}_{A}^{b)}\right].
\end{equation}
Lastly, with the replacement $\bar{\omega}\rightarrow i\frac{d}{dt}$ the first term is integrated. The gravitational radiation field is then determined by putting the two sources together in \eqref{eq:sourcev1gen}, and substituting this into \eqref{eq:radiated field}, to end up at the first sub-leading in BH velocities non-spinning correction to the Einstein quadrupole formula:
\begin{equation}\label{eq:amplitudes_scalar_v^1}
\begin{split}
h_{\text{TT},S^0}^{(1)\,ij}(T_{R},R,\boldsymbol{N},\boldsymbol{z}_1,\boldsymbol{z}_2)&=-\frac{\kappa^2 m_1}{8\pi R }\Pi^{ij}{}_{ab} N_l\Bigg[\frac{1}{2}
\left(\dot{\boldsymbol{v}}_{1}{\cdot}\boldsymbol{z}_{12}\delta^{lm}{-}\dot{v}_{1}^{m}z_{12}^{l}\right)\delta_{m}^{(a}(v_{1}+v_2)^{b)} {-}v_{1}^{l}v_{1}^{a}v_{1}^{b}
\\& 
-\frac{1}{2} \left(\frac{d}{dt}\left(\ddot{z}_{1}^{(a}z_{1}^{b)}(z_1+z_2)^l\right)+2(z_1+z_2)^lv_{1}^{(a}\dot{v}_{1}^{b)}\right)
\Bigg]+(1\leftrightarrow 2).
\end{split}
\end{equation}
Based on our derivation, this result is valid for generic \textit{closed} orbits, provided the corresponding EoM. However, the form of this waveform is different from the compact classical result in \eqref{eq:h1_v1_scalar}. This is not surprising since, as illustrated above, there is always the freedom of choice of IBP prescription, which casts the waveform into different forms. Finding the prescription, for which both the amplitude's and the classical waveforms match, could be cumbersome for generic closed orbits. Therefore, we specialize to the quasi-circular setting \eqref{eq:circorbits}; In the latter, we find perfect agreement between   \eqref{eq:amplitudes_scalar_v^1} and \eqref{eq:h1_v1_scalar}. We close with a remark on the correspondence between the classical source multipole moments leading to the gravitational radiation via the multipolar post-Minkowskian approach, and our ansatz to compute the associate gravitational waves using spinning scattering amplitudes. We saw in \cref{sec:hijclassical}, the sub-leading order result \eqref{eq:hijcircularv1} is built from both $\mathcal{I}_{ijk}$ and $\mathcal{J}_{ij}$. While at leading order in the BHs' velocities (see \cref{sec:linear_spin_waveform}), there exists a certain one-to-one correspondence between the source multipole moments, at sub-leading orders in velocities, no trivial correspondence can be extracted from our results.

\subsection{Radiated gravitational wave energy flux} \label{sec:results}

In the previous sections, we showed explicitly that the radiated gravitational field, $h^\text{TT}_{ij}$, computed using a classical approach and utilizing a 5-pt spinning scattering amplitude, agree in the aligned spin, general (and quasi-circular) orbit setup at the considered orders in the velocity and spin expansions. These are the gravitational waves emitted at an instant in the binary's evolution. Information about the frequency dynamics of the radiation is contained in the emitted gauge invariant gravitational wave energy flux. The latter is ultimately responsible for the inspiral of the two BHs and for the characteristic increase in gravitational wave frequency towards the merger, therefore, a crucial ingredient for gravitational wave search strategies.

In this section, we derive the instantaneous gravitational wave energy flux $\mathcal{F}$ using the TT metric perturbations at null infinity computed in the previous subsections to the respective orders in the spin and velocity expansions. In general, the total instantaneous energy loss $\mathcal{F}$ can be obtained with
\begin{align}
    \mathcal{F}=\frac{R^2}{32\pi}\int_{S^2} d\Omega \ \dot{h}^\text{TT}_{ij} \dot{h}^{\text{TT},ij}.
    \label{eq:generalfluxexpression}
\end{align}
Let us   return here to the justification for the replacements and claims made in \cref{sec:quadratic in spin}. The time dependence of $h^{\text{TT}}_{ij}$ is solely contained in the center of mass variables $n^a$ and $\lambda^a$, which, in the center of mass frame and for circular orbits, are related by $\frac{d}{dt}(n^a n^b)=\frac{1}{2}\frac{d}{dt}(n^a n^b-\lambda^a \lambda^b)$. Since only the time derivative of the radiated field, $\dot{h}^\text{TT}_{ij}$, enters in \eqref{eq:generalfluxexpression}, this justifies the replacement $n^a n^b\rightarrow\frac{1}{2}(n^an^b-\lambda^a\lambda^b)$ made in \cref{sec:quadratic in spin} at the level of the radiated field. Furthermore, this also shows that the gauge invariant energy flux is, in fact, independent of the IBP prescription discussed in \cref{sec:quadratic in spin}. Therefore, the latter can be viewed as a manifestation of the gauge freedom in the emitted waveform. Indeed, this extends to the Newman-Penrose scalar $\Psi_4\sim \bar{m}^\mu \bar{m}^\nu \ddot{h}^\text{TT}_{\mu\nu}$ in an identical fashion. Exploiting this, the gravitational wave energy flux is obtained by combining the scalar, \eqref{eq:einstein-quadrupole}, and linear-in-spin, \eqref{eq:amphijS1}, metric perturbations $h^{\text{TT}}_{ij}$ at leading order in the BHs' velocities, in \eqref{eq:generalfluxexpression}. For quasi-circular  orbits \eqref{eq:consvsol}, together with \eqref{eq:identityOmegaInt}, we find the energy loss
\begin{align}
\mathcal{F}^{(0)}_\text{circular}=\frac{32}{5}\frac{\mu^2x^5}{M^2}+\frac{2}{5}\frac{\mu^2 x^7}{M^2}(32a_+^2+a_-^2)+ \mathcal{O}(a^3_{1,2}, a_1 a_2).
\label{eq:circorbflux}
\end{align}
Recall from above that $x=(M\omega)^{2/3}$. This matches perfectly with the results reported in Refs.~\cite{Poisson:1997ha, Kidder:1992fr, Kidder:1995zr, Buonanno:2012rv, Marsat:2014xea, Siemonsen:2017yux} to the respective orders in spin. In addition to this match at leading order in the black holes velocities, the metric perturbations computed in \eqref{eq:amplitudes_scalar_v^1} and \eqref{eq:h1_v1_scalar} specialized to circular orbits reproduce the correct no-spin gravitational wave energy flux $\mathcal{F}^{(1)}_\text{circular}=0$ at the first sub-leading order in velocities; this is, again, consistent with the leading no-spin PN gravitational wave power (see, e.g., \cite{Kidder:1995zr}). Notice, we explicitly computed the quadratic-in-spin contributions only for one BH with spin: $S_1\neq 0, S_2\rightarrow 0$. However, as noted above, the classical derivation in \cref{sec:hijclassical} revealed that the high-order-in-spin contributions to the circular orbit $h^\text{TT}_{ij}$ emerge solely from the solution to the EoM, indicating that \eqref{eq:circorbflux} already contains the $a_1 a_2$-type interactions; this is the case, as can be seen in, for instance, \cite{Marsat:2014xea, Siemonsen:2017yux}, or from using \eqref{eq:hijeasier} together with \eqref{eq:generalfluxexpression}. At the level of the transverse traceless metric perturbations $h^\text{TT}_{ij}$, the classical derivation showed that \eqref{eq:classicalhijgeneralorbit} contains the complete all orders-in-spin information at leading order in velocities, since the remaining contributions to the radiation field -- i.e. $h_{ij,S^{\ell\ge 2}}^{(0)\text{TT}}=0$ -- vanish, \textit{without} a specific solution to the EoM. In the scattering amplitudes setting, we confirmed this explicitly up to $\ell=1$, since \eqref{eq:einstein-quadrupole} and \eqref{eq:amphijS1} agree with \eqref{eq:classicalhijgeneralorbit} (exploiting \eqref{eq:identityTT}), and we showed the necessary cancellation for $\ell=2$ in \cref{sec:quadratic in spin}. Therefore, we conjecture such cancellations to occur at arbitrary order in spin, such that the solution to the EoM provides the remaining spin-information, at leading order in velocities. The complete all-orders in spin gravitational power result partially presented in \eqref{eq:circorbflux} was determined in \cite{Siemonsen:2017yux}.

\section{Discussion }\label{sec:discusion}

We studied the relationship between the radiative dynamics of an aligned-spin spinning binary black hole from both, a classical, and a scattering amplitude perspective. For the former we employed the  multipolar post-Minkowskian formalism, whereas for the latter we proposed a dictionary  built from the 5-pt QFT scattering amplitude. More precisely, the dictionary maps the  classical limit of the 5-pt scattering amplitude of two massive spinning particles exchanging and emitting a graviton, to   the source entering in Einstein's equation.  Furthermore, we included information of the conservative dynamics using the classical equations of motion. We worked in linearized gravity, i.e., at tree-level, and to leading order in the black holes' velocities, but to all orders in their spin, as well as present preliminary results at sub-leading orders in velocities (in the no-spin limit). To leading order in the system's velocities, we showed that there exist a one-to-one correspondence between the source's multipole moments, and the scattering amplitudes. That is, the mass quadrupole in \eqref{eq:multipolemomentv0} corresponds to the scalar amplitude  \eqref{eq:scalar-amplitude}, while similarly, the current quadrupole in \eqref{eq:multipolemomentv0} is associated with the linear-in-spin amplitude \eqref{eq:linear_spin_v0}. This correspondence was made explicit  in the computation  of the  transverse-traceless part of the linear metric perturbations emitted to null infinity, as well as on the  gauge invariant gravitational wave energy flux. The latter agrees for quasi-circular orbits with the existing literature \cite{Kidder:1992fr, Kidder:1995zr, Marsat:2014xea, Siemonsen:2017yux}, both at the considered leading and sub-leading orders in the black holes' velocities. Therefore, gravitational waveforms and gauge invariant powers needed for detecting gravitational waves from \textit{inspiraling} black holes can be consistently computed from the classical limit of quantum \textit{scattering} amplitudes. 

The gravitational waveform is, in general, a gauge-dependent object, which makes a comparison between the classical and the scattering amplitude's derivations potentially difficult. In particular, and especially for general orbits and with spin effects, finding the corresponding gauge to undertake such comparisons can become cumbersome. In this work, we found evidence that such gauge freedom is related to the integration procedure used in the source for Einstein's equation, within the scattering amplitudes derivation. We demonstrated this explicitly for quasi-circular orbits, as this restriction simplifies the problem drastically. Importantly, we find that while the form of the gravitational radiation field is dependent upon the integration procedure used, the gauge-invariant gravitational wave power is independent of such a prescription -- as desired.

In this work, we focused entirely on the derivation of radiative degrees of freedom from the 5-pt scattering amplitude, while using a classical derivation of the system's conservative equations of motion. However, the conservative sector can efficiently be solved also, utilizing 4-pt scattering amplitudes. Therefore, one can envision combining 4-pt and 5-pt scattering amplitudes in such a way, as to remove the need for explicitly supplying the classical equations of motion. This might provide a pathway to deriving a gauge-invariant definition of the black holes' worldlines from the 5-point scattering amplitude in \eqref{eq:source_some1} that is entirely self-contained. In fact, for scalar sources, it was observed in Ref.~\cite{HariDass:1980tq} that the 4-pt graviton exchange amplitude satisfies the bodies' classical equations of motion suggesting such a relation. We leave exploring this avenue to future work. 

The amplitudes-based construction of the radiated field  \eqref{eq:radiated field}, provided in this work, has implicitly  used  the on-shell condition for the outgoing massive particles $\delta(p_i{\cdot}q_i)$, which discards terms quadratic in the velocities as indicated by the quantum corrections to the particles trajectories  $z_{Q}(\tau_i)$ in  \eqref{eq:radiation_kernell_gen}. These corrections can become important if convoluted with superclassical terms coming from loop amplitudes. This then   hints that at higher orders in perturbation theory, a  subtraction scheme would be needed to cancel those superclassical contributions  at the level of the gauge invariant observable, which in  this case corresponds to  the radiated energy flux  $\mathcal{F} \sim \int d\omega \dot{h}^{ij}\dot{h}_{ij}$; in addition, it would be desirable to study the connection of our approach and that of analytic continuation methods of scattering observables \cite{Kalin:2019inp, Kalin:2019rwq,Herrmann:2021lqe} .

 Besides,  exploring gauge fixing procedures that allow to match the general orbit result \eqref{eq:amplitudes_scalar_v^1} to the  classical result \eqref{eq:h1_v1_scalar}, as well as the inclusion of spin effects at sub-leading order in velocity is left for future work. Furthermore, in the context of scattering amplitudes, higher orders in velocities are naturally included. However, for closed orbits, these corrections are consistent only -- by virtue of the  virial theorem -- when also higher orders in the gravitational constant $G$ are considered. For instance, at quadratic order in the BHs' velocities, the radiated field could contain contributions from both the tree-level and the one-loop 5-pt scattering amplitudes. One might wonder whether the amplitudes approach could reproduce the higher-order corrections to the energy flux for non-spinning binary black holes \cite{Blanchet:2013haa}.  

Finally, the source \eqref{eq:source_some1} was written in the Born approximation, where the initial state consists of two particles in their plane-wave states. However, the long-range nature of the gravitational  interactions renders the Born approximation to be invalid in this setting. Although this is expected to be a higher-$G$-effect (or equivalently a higher-$v$-effect in the closed orbit case), it plays an important role in the determination of the correct gravitational waveform. A modification to the Born approximation was proposed in \cite{HariDass:1980tq}, and claimed to contain all non-perturbative aspects of the S-matrix elements. We leave the exploration of this proposal for future work.

\acknowledgments

We would like to thank Alfredo Guevara for his collaboration during the initial stages of this work, and continuous encouragements throughout. We are also grateful for enlightening discussions with Chris Kavanagh, Alok Laddha, Gustav Mogull, Donal O'Connell,  Jan Plefka, Shan-Ming Ruan, Jan Steinhoff, and Justin Vines. The authors acknowledge financial support by the Natural Sciences and Engineering Research Council of Canada (NSERC). Research at Perimeter Institute is supported in part by the Government of Canada through the Department of Innovation, Science and Economic Development Canada and by the Province of Ontario through the Ministry of Economic Development, Job Creation and Trade. This research was undertaken thanks in part to funding from the Canada First Research Excellence Fund through the Arthur B. McDonald Canadian Astroparticle Physics Research Institute.

\appendix

\section{Useful integrals and identities}\label{ap:integrals}
Here we write out the identity used in \cref{sec:linear_spin_waveform} for the comparison of the gravitational waveforms at linear order in spin. That is, given two vectors $a^i$ and $b^i$, and the TT projector defined in  \eqref{eq:TTprojector}, we have  \cite{Kidder:1995zr, Buonanno:2012rv}
\begin{align}
\Pi^{ab}{}_{ij}b^j\varepsilon^{i}{}_{k\ell}a^k N^\ell  =\Pi^{ab}{}_{ij} a^{i}\varepsilon^{j}{}_{k\ell}b^k N^\ell .
\label{eq:identityTT}
\end{align}
Furthermore, the following identity  \cite{Maggiore:1900zz}, was used in the computation of the energy flux in \cref{sec:results}
\begin{align}
\int_{S^2} d\Omega \ N_{i_1 \dots i_{2\ell}} =\frac{4\pi}{(2\ell+1)!!}(\delta_{i_1 i_2}\delta_{i_3 i_4}\dots \delta_{i_{2\ell-1} i_{2\ell}}+\dots).
\label{eq:identityOmegaInt}
\end{align}
In addition, the following integrals were used during the computation of gravitational radiation from the amplitudes perspective:
\begin{equation}
\begin{split}
\int\frac{d^{3}q}{(2\pi)^{3}}e^{i\bs{q}{\cdot}\bs{z}}\frac{1}{\bs{q}^{2}}&=\frac{1}{4\pi|\bs{z}|},\\
\int\frac{d^{3}q}{(2\pi)^{3}}e^{i\bs{q}{\cdot}\bs{z}}\frac{q^{i}}{\bs{q}^{2}}&=\frac{iz^{i}}{4\pi|\bs{z}|^{3}},\\
\int\frac{d^{3}q}{(2\pi)^{3}}e^{i\bs{q}{\cdot}\bs{z}}\frac{q^{i}q^{j}}{\bs{q}^{2}}&=\frac{1}{4\pi|\bs{z}|^{5}}\left[|\bs{z}|^{2}\delta^{ij}-3z^{i}z^{j}\right],\\
\int\frac{d^{3}q}{(2\pi)^{3}}e^{i\bs{q}{\cdot}\bs{z}}\frac{q^{i}q^{j}}{\bs{q}^{4}}&=\frac{1}{8\pi|\bs{z}|^{3}}\left[|\bs{z}|^{2}\delta^{ij}-z^{i}z^{j}\right],\\
\int\frac{d^{3}q}{(2\pi)^{3}}e^{i\bs{q}{\cdot}\bs{z}}\frac{q^{i}q^{j}q^{k}}{\bs{q}^{4}}&=-\frac{i}{8\pi|\bs{z}|^{5}}\left[|\bs{z}|^{2}\left(z^{i}\delta^{jk}+z^{j}\delta^{ik}+z^{k}\delta^{ij}\right)-3z^{i}z^{j}z^{k}\right].
\end{split}\label{eq:integrals}
\end{equation}

\section{The quadratic in spin EoM}\label{sec: eom quad spin}

In this appendix, we expand the classical equations of motion in \eqref{eq:EoM} to quadratic order in the black holes' spins (used in  \cref{sec:quadratic in spin}). After setting $S_2=0$, and expanding to second  order in $S_1$,  as well as  taking the leading order in velocity, the equation of motion reduce to
\begin{equation}
    \dot{v}_1^l =  \frac{-m_2\kappa^2}{32\pi }\left[  \frac{z_{12}^l}{r^3}+\frac{1}{2m_1^2}S_1^iS_1^j\partial^l\partial_i\partial_j\frac{1}{r}\right].
    \label{eq:eom quad spin}
\end{equation}
The spatial derivatives acting on $1/r$ result in contractions of a symmetric trace-free tensor
\begin{equation}
\partial^l\partial_i\partial_j\frac{1}{r} = \frac{3}{r^5}\left[\delta_{ij}z_{12}^l+2\delta^l_{(i}z_{12,j)}-5\frac{z_{12,i}z_{12,j}z_{12}^l}{r^2}\right].
\label{eq:spatialderivatives}
\end{equation}
Furthermore, we use these equations recursively, to remove powers of $1/r$\footnote{By restoring Newton's constant $G$, the equations of motion can be used to remove powers of $G$ in the numerator.}. Since we are interested in the quadratic-in-spin contribution only, we consider only the scalar part of  \eqref{eq:eom quad spin} (as well as the analogous equation for $v_2^i$) to rewrite \eqref{eq:spatialderivatives} as follows
\begin{equation}
    \partial^l\partial_i\partial_j\frac{1}{r} \rightarrow 
    -\frac{32\pi }{\kappa^2}\frac{3}{2m_2r^2}\left[\left(\delta_{ij}{-}\frac{5z_{12,i}z_{12,j}}{r^2}\right)\left(\dot{v}_1^l {-}\frac{m_2}{m_1}\dot{v}_2^l  \right)+2 \delta^l_{(i}\left(\dot{v}_{1,j)}{-}\frac{m_2}{m_1} \dot{v}_{2,j)}  \right) \right] {+}\mathcal{O}(S_1^2),
\end{equation}
Notice a factor of $1/3$ arises from symmetrization. 
This, then finally allows us to write the quadratic-in-spin equations of motion as
\begin{equation}
     \dot{v}_1^l = {-}\frac{m_2\kappa^2}{32\pi}  \frac{z_{12}^l}{r^3}{+}\frac{3}{4}\frac{S_1^iS_1^j}{m_1^2r^2}\left[\left(\delta_{ij}{-}\frac{5z_{12,i}z_{12,j}}{r^2}\right)\left(\dot{v}_1^l {-}\frac{m_2}{m_1}\dot{v}_2^l  \right){+}2 \delta^l_{(i}\left(\dot{v}_{1,j)}{-}\frac{m_2}{m_1} \dot{v}_{2,j)}  \right) \right].
\end{equation}
And analogously we also find
\begin{equation}
     \dot{v}_2^l = \frac{m_1\kappa^2}{32\pi }  \frac{z_{12}^l}{r^3}+\frac{3}{4}\frac{S_1^iS_1^j}{m_1^2r^2}\left[\left(\delta_{ij}-\frac{5z_{12,i}z_{12,j}}{r^2}\right)\left(\dot{v}_2^l -\frac{m_1}{m_2}\dot{v}_1^l  \right)+2 \delta^l_{(i}\left(\dot{v}_{2,j)}-\frac{m_1}{m_2} \dot{v}_{1,j)}  \right) \right].
\end{equation}

\bibliographystyle{JHEP}

\bibliography{bib.bib}

\providecommand{\href}[2]{#2}\begingroup\raggedright\begin{thebibliography}{100}

\bibitem{1918SPAW154E}
A.~{Einstein}, \emph{{{\"U}ber Gravitationswellen}}, {\emph{Sitzungsberichte
  der K{\"o}niglich Preu{\ss}ischen Akademie der Wissenschaften (Berlin} (1918)
  154}.

\bibitem{LIGOScientific:2016aoc}
{\scshape LIGO Scientific, Virgo} collaboration, \emph{{Observation of
  Gravitational Waves from a Binary Black Hole Merger}},
  \href{https://doi.org/10.1103/PhysRevLett.116.061102}{\emph{Phys. Rev. Lett.}
  {\bfseries 116} (2016) 061102}
  [\href{https://arxiv.org/abs/1602.03837}{{\ttfamily 1602.03837}}].

\bibitem{Blanchet:2013haa}
L.~Blanchet, \emph{{Gravitational Radiation from Post-Newtonian Sources and
  Inspiralling Compact Binaries}},
  \href{https://doi.org/10.12942/lrr-2014-2}{\emph{Living Rev. Rel.} {\bfseries
  17} (2014) 2} [\href{https://arxiv.org/abs/1310.1528}{{\ttfamily
  1310.1528}}].

\bibitem{Futamase:2007zz}
T.~Futamase and Y.~Itoh, \emph{{The post-Newtonian approximation for
  relativistic compact binaries}},
  \href{https://doi.org/10.12942/lrr-2007-2}{\emph{Living Rev. Rel.} {\bfseries
  10} (2007) 2}.

\bibitem{Pretorius:2005gq}
F.~Pretorius, \emph{{Evolution of binary black hole spacetimes}},
  \href{https://doi.org/10.1103/PhysRevLett.95.121101}{\emph{Phys. Rev. Lett.}
  {\bfseries 95} (2005) 121101}
  [\href{https://arxiv.org/abs/gr-qc/0507014}{{\ttfamily gr-qc/0507014}}].

\bibitem{1973Teukolsky}
S.~A. {Teukolsky}, \emph{{Perturbations of a Rotating Black Hole. I.
  Fundamental Equations for Gravitational, Electromagnetic, and Neutrino-Field
  Perturbations}}, \href{https://doi.org/10.1086/152444}{\emph{Apj} {\bfseries
  185} (1973) 635}.

\bibitem{Kokkotas:1999bd}
K.~D. Kokkotas and B.~G. Schmidt, \emph{{Quasinormal modes of stars and black
  holes}}, \href{https://doi.org/10.12942/lrr-1999-2}{\emph{Living Rev. Rel.}
  {\bfseries 2} (1999) 2}
  [\href{https://arxiv.org/abs/gr-qc/9909058}{{\ttfamily gr-qc/9909058}}].

\bibitem{PhysRevD.59.084006}
A.~Buonanno and T.~Damour, \emph{Effective one-body approach to general
  relativistic two-body dynamics},
  \href{https://doi.org/10.1103/PhysRevD.59.084006}{\emph{Phys. Rev. D}
  {\bfseries 59} (1999) 084006}.

\bibitem{Buonanno:2000ef}
A.~Buonanno and T.~Damour, \emph{{Transition from inspiral to plunge in binary
  black hole coalescences}},
  \href{https://doi.org/10.1103/PhysRevD.62.064015}{\emph{Phys. Rev.}
  {\bfseries D62} (2000) 064015}
  [\href{https://arxiv.org/abs/gr-qc/0001013}{{\ttfamily gr-qc/0001013}}].

\bibitem{Santamaria:2010yb}
L.~Santamaria et~al., \emph{{Matching post-Newtonian and numerical relativity
  waveforms: systematic errors and a new phenomenological model for
  non-precessing black hole binaries}},
  \href{https://doi.org/10.1103/PhysRevD.82.064016}{\emph{Phys. Rev. D}
  {\bfseries 82} (2010) 064016}
  [\href{https://arxiv.org/abs/1005.3306}{{\ttfamily 1005.3306}}].

\bibitem{Damour_2016}
T.~Damour, \emph{Gravitational scattering, post-minkowskian approximation, and
  effective-one-body theory},
  \href{https://doi.org/10.1103/physrevd.94.104015}{\emph{Physical Review D}
  {\bfseries 94} (2016) }.

\bibitem{Porto:2016pyg}
R.~A. Porto, \emph{{The effective field theorist’s approach to gravitational
  dynamics}}, \href{https://doi.org/10.1016/j.physrep.2016.04.003}{\emph{Phys.
  Rept.} {\bfseries 633} (2016) 1}
  [\href{https://arxiv.org/abs/1601.04914}{{\ttfamily 1601.04914}}].

\bibitem{Goldberger:2017vcg}
W.~D. Goldberger and A.~K. Ridgway, \emph{{Bound states and the classical
  double copy}}, \href{https://doi.org/10.1103/PhysRevD.97.085019}{\emph{Phys.
  Rev. D} {\bfseries 97} (2018) 085019}
  [\href{https://arxiv.org/abs/1711.09493}{{\ttfamily 1711.09493}}].

\bibitem{Goldberger:2017ogt}
W.~D. Goldberger, J.~Li and S.~G. Prabhu, \emph{{Spinning particles, axion
  radiation, and the classical double copy}},
  \href{https://doi.org/10.1103/PhysRevD.97.105018}{\emph{Phys. Rev.}
  {\bfseries D97} (2018) 105018}
  [\href{https://arxiv.org/abs/1712.09250}{{\ttfamily 1712.09250}}].

\bibitem{Vines:2018gqi}
J.~Vines, J.~Steinhoff and A.~Buonanno, \emph{{Spinning-black-hole scattering
  and the test-black-hole limit at second post-Minkowskian order}},
  \href{https://doi.org/10.1103/PhysRevD.99.064054}{\emph{Phys. Rev. D}
  {\bfseries 99} (2019) 064054}
  [\href{https://arxiv.org/abs/1812.00956}{{\ttfamily 1812.00956}}].

\bibitem{Damour:2019lcq}
T.~Damour, \emph{{Classical and quantum scattering in post-Minkowskian
  gravity}}, \href{https://doi.org/10.1103/PhysRevD.102.024060}{\emph{Phys.
  Rev. D} {\bfseries 102} (2020) 024060}
  [\href{https://arxiv.org/abs/1912.02139}{{\ttfamily 1912.02139}}].

\bibitem{Damour:2020tta}
T.~Damour, \emph{{Radiative contribution to classical gravitational scattering
  at the third order in $G$}},
  \href{https://doi.org/10.1103/PhysRevD.102.124008}{\emph{Phys. Rev. D}
  {\bfseries 102} (2020) 124008}
  [\href{https://arxiv.org/abs/2010.01641}{{\ttfamily 2010.01641}}].

\bibitem{Kalin:2019rwq}
G.~K\"alin and R.~A. Porto, \emph{{From Boundary Data to Bound States}},
  \href{https://doi.org/10.1007/JHEP01(2020)072}{\emph{JHEP} {\bfseries 01}
  (2020) 072} [\href{https://arxiv.org/abs/1910.03008}{{\ttfamily
  1910.03008}}].

\bibitem{Kalin:2019inp}
G.~K\"alin and R.~A. Porto, \emph{{From boundary data to bound states. Part II.
  Scattering angle to dynamical invariants (with twist)}},
  \href{https://doi.org/10.1007/JHEP02(2020)120}{\emph{JHEP} {\bfseries 02}
  (2020) 120} [\href{https://arxiv.org/abs/1911.09130}{{\ttfamily
  1911.09130}}].

\bibitem{Kalin:2020mvi}
G.~K\"alin and R.~A. Porto, \emph{{Post-Minkowskian Effective Field Theory for
  Conservative Binary Dynamics}},
  \href{https://doi.org/10.1007/JHEP11(2020)106}{\emph{JHEP} {\bfseries 11}
  (2020) 106} [\href{https://arxiv.org/abs/2006.01184}{{\ttfamily
  2006.01184}}].

\bibitem{Kalin:2020fhe}
G.~K\"alin, Z.~Liu and R.~A. Porto, \emph{{Conservative Dynamics of Binary
  Systems to Third Post-Minkowskian Order from the Effective Field Theory
  Approach}}, \href{https://doi.org/10.1103/PhysRevLett.125.261103}{\emph{Phys.
  Rev. Lett.} {\bfseries 125} (2020) 261103}
  [\href{https://arxiv.org/abs/2007.04977}{{\ttfamily 2007.04977}}].

\bibitem{Goldberger:2020fot}
W.~D. Goldberger, J.~Li and I.~Z. Rothstein, \emph{{Non-conservative effects on
  Spinning Black Holes from World-Line Effective Field Theory}},
  \href{https://arxiv.org/abs/2012.14869}{{\ttfamily 2012.14869}}.

\bibitem{Brandhuber:2021kpo}
A.~Brandhuber, G.~Chen, G.~Travaglini and C.~Wen, \emph{{A new gauge-invariant
  double copy for heavy-mass effective theory}},
  \href{https://doi.org/10.1007/JHEP07(2021)047}{\emph{JHEP} {\bfseries 07}
  (2021) 047} [\href{https://arxiv.org/abs/2104.11206}{{\ttfamily
  2104.11206}}].

\bibitem{Brandhuber:2021eyq}
A.~Brandhuber, G.~Chen, G.~Travaglini and C.~Wen, \emph{{Classical
  gravitational scattering from a gauge-invariant double copy}},
  \href{https://doi.org/10.1007/JHEP10(2021)118}{\emph{JHEP} {\bfseries 10}
  (2021) 118} [\href{https://arxiv.org/abs/2108.04216}{{\ttfamily
  2108.04216}}].

\bibitem{Cheung:2020gyp}
C.~Cheung and M.~P. Solon, \emph{{Classical gravitational scattering at $
  \mathcal{O} $(G$^{3}$) from Feynman diagrams}},
  \href{https://doi.org/10.1007/JHEP06(2020)144}{\emph{JHEP} {\bfseries 06}
  (2020) 144} [\href{https://arxiv.org/abs/2003.08351}{{\ttfamily
  2003.08351}}].

\bibitem{Cheung:2018wkq}
C.~Cheung, I.~Z. Rothstein and M.~P. Solon, \emph{{From Scattering Amplitudes
  to Classical Potentials in the Post-Minkowskian Expansion}},
  \href{https://doi.org/10.1103/PhysRevLett.121.251101}{\emph{Phys. Rev. Lett.}
  {\bfseries 121} (2018) 251101}
  [\href{https://arxiv.org/abs/1808.02489}{{\ttfamily 1808.02489}}].

\bibitem{Bern:2019nnu}
Z.~Bern, C.~Cheung, R.~Roiban, C.-H. Shen, M.~P. Solon and M.~Zeng,
  \emph{{Scattering Amplitudes and the Conservative Hamiltonian for Binary
  Systems at Third Post-Minkowskian Order}},
  \href{https://doi.org/10.1103/PhysRevLett.122.201603}{\emph{Phys. Rev. Lett.}
  {\bfseries 122} (2019) 201603}
  [\href{https://arxiv.org/abs/1901.04424}{{\ttfamily 1901.04424}}].

\bibitem{Bern:2019crd}
Z.~Bern, C.~Cheung, R.~Roiban, C.-H. Shen, M.~P. Solon and M.~Zeng,
  \emph{{Black Hole Binary Dynamics from the Double Copy and Effective
  Theory}}, \href{https://doi.org/10.1007/JHEP10(2019)206}{\emph{JHEP}
  {\bfseries 10} (2019) 206}
  [\href{https://arxiv.org/abs/1908.01493}{{\ttfamily 1908.01493}}].

\bibitem{Bern:2021dqo}
Z.~Bern, J.~Parra-Martinez, R.~Roiban, M.~S. Ruf, C.-H. Shen, M.~P. Solon
  et~al., \emph{{Scattering Amplitudes and Conservative Binary Dynamics at
  ${\cal O}(G^4)$}},
  \href{https://doi.org/10.1103/PhysRevLett.126.171601}{\emph{Phys. Rev. Lett.}
  {\bfseries 126} (2021) 171601}
  [\href{https://arxiv.org/abs/2101.07254}{{\ttfamily 2101.07254}}].

\bibitem{Bjerrum-Bohr:2018xdl}
N.~E.~J. Bjerrum-Bohr, P.~H. Damgaard, G.~Festuccia, L.~Plant\'e and
  P.~Vanhove, \emph{{General Relativity from Scattering Amplitudes}},
  \href{https://doi.org/10.1103/PhysRevLett.121.171601}{\emph{Phys. Rev. Lett.}
  {\bfseries 121} (2018) 171601}
  [\href{https://arxiv.org/abs/1806.04920}{{\ttfamily 1806.04920}}].

\bibitem{Cristofoli:2019neg}
A.~Cristofoli, N.~E.~J. Bjerrum-Bohr, P.~H. Damgaard and P.~Vanhove,
  \emph{{Post-Minkowskian Hamiltonians in general relativity}},
  \href{https://doi.org/10.1103/PhysRevD.100.084040}{\emph{Phys. Rev. D}
  {\bfseries 100} (2019) 084040}
  [\href{https://arxiv.org/abs/1906.01579}{{\ttfamily 1906.01579}}].

\bibitem{Bjerrum-Bohr:2019kec}
N.~E.~J. Bjerrum-Bohr, A.~Cristofoli and P.~H. Damgaard,
  \emph{{Post-Minkowskian Scattering Angle in Einstein Gravity}},
  \href{https://doi.org/10.1007/JHEP08(2020)038}{\emph{JHEP} {\bfseries 08}
  (2020) 038} [\href{https://arxiv.org/abs/1910.09366}{{\ttfamily
  1910.09366}}].

\bibitem{Bjerrum-Bohr:2021vuf}
N.~E.~J. Bjerrum-Bohr, P.~H. Damgaard, L.~Plant\'e and P.~Vanhove,
  \emph{{Classical Gravity from Loop Amplitudes}},
  \href{https://arxiv.org/abs/2104.04510}{{\ttfamily 2104.04510}}.

\bibitem{DiVecchia:2020ymx}
P.~Di~Vecchia, C.~Heissenberg, R.~Russo and G.~Veneziano, \emph{{Universality
  of ultra-relativistic gravitational scattering}},
  \href{https://doi.org/10.1016/j.physletb.2020.135924}{\emph{Phys. Lett. B}
  {\bfseries 811} (2020) 135924}
  [\href{https://arxiv.org/abs/2008.12743}{{\ttfamily 2008.12743}}].

\bibitem{DiVecchia:2021ndb}
P.~Di~Vecchia, C.~Heissenberg, R.~Russo and G.~Veneziano, \emph{{Radiation
  Reaction from Soft Theorems}},
  \href{https://doi.org/10.1016/j.physletb.2021.136379}{\emph{Phys. Lett. B}
  {\bfseries 818} (2021) 136379}
  [\href{https://arxiv.org/abs/2101.05772}{{\ttfamily 2101.05772}}].

\bibitem{Bern:2020buy}
Z.~Bern, A.~Luna, R.~Roiban, C.-H. Shen and M.~Zeng, \emph{{Spinning black hole
  binary dynamics, scattering amplitudes, and effective field theory}},
  \href{https://doi.org/10.1103/PhysRevD.104.065014}{\emph{Phys. Rev. D}
  {\bfseries 104} (2021) 065014}
  [\href{https://arxiv.org/abs/2005.03071}{{\ttfamily 2005.03071}}].

\bibitem{Chung:2019duq}
M.-Z. Chung, Y.-T. Huang and J.-W. Kim, \emph{{Classical potential for general
  spinning bodies}}, \href{https://doi.org/10.1007/JHEP09(2020)074}{\emph{JHEP}
  {\bfseries 09} (2020) 074}
  [\href{https://arxiv.org/abs/1908.08463}{{\ttfamily 1908.08463}}].

\bibitem{Chung:2020rrz}
M.-Z. Chung, Y.-t. Huang, J.-W. Kim and S.~Lee, \emph{{Complete Hamiltonian for
  spinning binary systems at first post-Minkowskian order}},
  \href{https://doi.org/10.1007/JHEP05(2020)105}{\emph{JHEP} {\bfseries 05}
  (2020) 105} [\href{https://arxiv.org/abs/2003.06600}{{\ttfamily
  2003.06600}}].

\bibitem{Cachazo:2017jef}
F.~Cachazo and A.~Guevara, \emph{{Leading Singularities and Classical
  Gravitational Scattering}},
  \href{https://doi.org/10.1007/JHEP02(2020)181}{\emph{JHEP} {\bfseries 02}
  (2020) 181} [\href{https://arxiv.org/abs/1705.10262}{{\ttfamily
  1705.10262}}].

\bibitem{Guevara:2017csg}
A.~Guevara, \emph{{Holomorphic Classical Limit for Spin Effects in
  Gravitational and Electromagnetic Scattering}},
  \href{https://doi.org/10.1007/JHEP04(2019)033}{\emph{JHEP} {\bfseries 04}
  (2019) 033} [\href{https://arxiv.org/abs/1706.02314}{{\ttfamily
  1706.02314}}].

\bibitem{Guevara:2018wpp}
A.~Guevara, A.~Ochirov and J.~Vines, \emph{{Scattering of Spinning Black Holes
  from Exponentiated Soft Factors}},
  \href{https://doi.org/10.1007/JHEP09(2019)056}{\emph{JHEP} {\bfseries 09}
  (2019) 056} [\href{https://arxiv.org/abs/1812.06895}{{\ttfamily
  1812.06895}}].

\bibitem{Guevara:2019fsj}
A.~Guevara, A.~Ochirov and J.~Vines, \emph{{Black-hole scattering with general
  spin directions from minimal-coupling amplitudes}},
  \href{https://arxiv.org/abs/1906.10071}{{\ttfamily 1906.10071}}.

\bibitem{Aoude:2020onz}
R.~Aoude, K.~Haddad and A.~Helset, \emph{{On-shell heavy particle effective
  theories}}, \href{https://doi.org/10.1007/JHEP05(2020)051}{\emph{JHEP}
  {\bfseries 05} (2020) 051}
  [\href{https://arxiv.org/abs/2001.09164}{{\ttfamily 2001.09164}}].

\bibitem{Bautista:2019evw}
Y.~F. Bautista and A.~Guevara, \emph{{On the Double Copy for Spinning Matter}},
   \href{https://arxiv.org/abs/1908.11349}{{\ttfamily 1908.11349}}.

\bibitem{Blumlein:2020pyo}
J.~Bl\"umlein, A.~Maier, P.~Marquard and G.~Sch\"afer, \emph{{The fifth-order
  post-Newtonian Hamiltonian dynamics of two-body systems from an effective
  field theory approach: potential contributions}},
  \href{https://doi.org/10.1016/j.nuclphysb.2021.115352}{\emph{Nucl. Phys. B}
  {\bfseries 965} (2021) 115352}
  [\href{https://arxiv.org/abs/2010.13672}{{\ttfamily 2010.13672}}].

\bibitem{Blumlein:2020znm}
J.~Bl\"umlein, A.~Maier, P.~Marquard and G.~Sch\"afer, \emph{{Testing binary
  dynamics in gravity at the sixth post-Newtonian level}},
  \href{https://doi.org/10.1016/j.physletb.2020.135496}{\emph{Phys. Lett. B}
  {\bfseries 807} (2020) 135496}
  [\href{https://arxiv.org/abs/2003.07145}{{\ttfamily 2003.07145}}].

\bibitem{Laddha:2018rle}
A.~Laddha and A.~Sen, \emph{{Gravity Waves from Soft Theorem in General
  Dimensions}}, \href{https://doi.org/10.1007/JHEP09(2018)105}{\emph{JHEP}
  {\bfseries 09} (2018) 105}
  [\href{https://arxiv.org/abs/1801.07719}{{\ttfamily 1801.07719}}].

\bibitem{Saha:2019tub}
A.~P. Saha, B.~Sahoo and A.~Sen, \emph{{Proof of the classical soft graviton
  theorem in $D$ = 4}},
  \href{https://doi.org/10.1007/JHEP06(2020)153}{\emph{JHEP} {\bfseries 06}
  (2020) 153} [\href{https://arxiv.org/abs/1912.06413}{{\ttfamily
  1912.06413}}].

\bibitem{Ghosh:2021hsk}
D.~Ghosh and B.~Sahoo, \emph{{Spin Dependent Gravitational Tail Memory in
  $D=4$}},  \href{https://arxiv.org/abs/2106.10741}{{\ttfamily 2106.10741}}.

\bibitem{Kosower:2018adc}
D.~A. Kosower, B.~Maybee and D.~O'Connell, \emph{{Amplitudes, Observables, and
  Classical Scattering}},
  \href{https://doi.org/10.1007/JHEP02(2019)137}{\emph{JHEP} {\bfseries 02}
  (2019) 137} [\href{https://arxiv.org/abs/1811.10950}{{\ttfamily
  1811.10950}}].

\bibitem{Bautista:2019tdr}
Y.~F. Bautista and A.~Guevara, \emph{{From Scattering Amplitudes to Classical
  Physics: Universality, Double Copy and Soft Theorems}},
  \href{https://arxiv.org/abs/1903.12419}{{\ttfamily 1903.12419}}.

\bibitem{1987KPT}
V.~B. Braginsky and K.~S. Thorne, \emph{Gravitational-wave bursts with memory
  and experimental prospects}, .

\bibitem{Herrmann:2021lqe}
E.~Herrmann, J.~Parra-Martinez, M.~S. Ruf and M.~Zeng, \emph{{Gravitational
  Bremsstrahlung from Reverse Unitarity}},
  \href{https://doi.org/10.1103/PhysRevLett.126.201602}{\emph{Phys. Rev. Lett.}
  {\bfseries 126} (2021) 201602}
  [\href{https://arxiv.org/abs/2101.07255}{{\ttfamily 2101.07255}}].

\bibitem{Herrmann:2021tct}
E.~Herrmann, J.~Parra-Martinez, M.~S. Ruf and M.~Zeng, \emph{{Radiative
  Classical Gravitational Observables at $\mathcal O(G^3)$ from Scattering
  Amplitudes}},  \href{https://arxiv.org/abs/2104.03957}{{\ttfamily
  2104.03957}}.

\bibitem{DiVecchia:2021bdo}
P.~Di~Vecchia, C.~Heissenberg, R.~Russo and G.~Veneziano, \emph{{The eikonal
  approach to gravitational scattering and radiation at $ \mathcal{O}
  $(G$^{3}$)}}, \href{https://doi.org/10.1007/JHEP07(2021)169}{\emph{JHEP}
  {\bfseries 07} (2021) 169}
  [\href{https://arxiv.org/abs/2104.03256}{{\ttfamily 2104.03256}}].

\bibitem{Bini:2021gat}
D.~Bini, T.~Damour and A.~Geralico, \emph{{Radiative contributions to
  gravitational scattering}},
  \href{https://doi.org/10.1103/PhysRevD.104.084031}{\emph{Phys. Rev. D}
  {\bfseries 104} (2021) 084031}
  [\href{https://arxiv.org/abs/2107.08896}{{\ttfamily 2107.08896}}].

\bibitem{Riva:2021vnj}
M.~M. Riva and F.~Vernizzi, \emph{{Radiated momentum in the Post-Minkowskian
  worldline approach via reverse unitarity}},
  \href{https://arxiv.org/abs/2110.10140}{{\ttfamily 2110.10140}}.

\bibitem{Mogull:2020sak}
G.~Mogull, J.~Plefka and J.~Steinhoff, \emph{{Classical black hole scattering
  from a worldline quantum field theory}},
  \href{https://doi.org/10.1007/JHEP02(2021)048}{\emph{JHEP} {\bfseries 02}
  (2021) 048} [\href{https://arxiv.org/abs/2010.02865}{{\ttfamily
  2010.02865}}].

\bibitem{Jakobsen:2021smu}
G.~U. Jakobsen, G.~Mogull, J.~Plefka and J.~Steinhoff, \emph{{Classical
  Gravitational Bremsstrahlung from a Worldline Quantum Field Theory}},
  \href{https://doi.org/10.1103/PhysRevLett.126.201103}{\emph{Phys. Rev. Lett.}
  {\bfseries 126} (2021) 201103}
  [\href{https://arxiv.org/abs/2101.12688}{{\ttfamily 2101.12688}}].

\bibitem{Mougiakakos:2021ckm}
S.~Mougiakakos, M.~M. Riva and F.~Vernizzi, \emph{{Gravitational Bremsstrahlung
  in the post-Minkowskian effective field theory}},
  \href{https://doi.org/10.1103/PhysRevD.104.024041}{\emph{Phys. Rev. D}
  {\bfseries 104} (2021) 024041}
  [\href{https://arxiv.org/abs/2102.08339}{{\ttfamily 2102.08339}}].

\bibitem{Jakobsen:2021lvp}
G.~U. Jakobsen, G.~Mogull, J.~Plefka and J.~Steinhoff, \emph{{Gravitational
  Bremsstrahlung and Hidden Supersymmetry of Spinning Bodies}},
  \href{https://arxiv.org/abs/2106.10256}{{\ttfamily 2106.10256}}.

\bibitem{Cristofoli:2021vyo}
A.~Cristofoli, R.~Gonzo, D.~A. Kosower and D.~O'Connell, \emph{{Waveforms from
  Amplitudes}},  \href{https://arxiv.org/abs/2107.10193}{{\ttfamily
  2107.10193}}.

\bibitem{Kol:2021jjc}
U.~Kol, D.~O'connell and O.~Telem, \emph{{The Radial Action from Probe
  Amplitudes to All Orders}},
  \href{https://arxiv.org/abs/2109.12092}{{\ttfamily 2109.12092}}.

\bibitem{Bautista:2021wfy}
Y.~F. Bautista, A.~Guevara, C.~Kavanagh and J.~Vines, \emph{{From Scattering in
  Black Hole Backgrounds to Higher-Spin Amplitudes: Part I}},
  \href{https://arxiv.org/abs/2107.10179}{{\ttfamily 2107.10179}}.

\bibitem{BCGV}
Y.~F. Bautista, C.~Kavanagh, A.~Guevara and J.~Vines, \emph{{From Scattering in
  Black Hole Backgrounds to Higher-Spin Amplitudes: Part II. In preparation}},
  .

\bibitem{Falkowski:2020aso}
A.~Falkowski and C.~S. Machado, \emph{{Soft Matters, or the Recursions with
  Massive Spinors}}, \href{https://doi.org/10.1007/JHEP05(2021)238}{\emph{JHEP}
  {\bfseries 05} (2021) 238}
  [\href{https://arxiv.org/abs/2005.08981}{{\ttfamily 2005.08981}}].

\bibitem{Chiodaroli:2021eug}
M.~Chiodaroli, H.~Johansson and P.~Pichini, \emph{{Compton Black-Hole
  Scattering for $s \leq 5/2$}},
  \href{https://arxiv.org/abs/2107.14779}{{\ttfamily 2107.14779}}.

\bibitem{Dlapa:2021npj}
C.~Dlapa, G.~K\"alin, Z.~Liu and R.~A. Porto, \emph{{Dynamics of Binary Systems
  to Fourth Post-Minkowskian Order from the Effective Field Theory Approach}},
  \href{https://arxiv.org/abs/2106.08276}{{\ttfamily 2106.08276}}.

\bibitem{Saketh:2021sri}
M.~V.~S. Saketh, J.~Vines, J.~Steinhoff and A.~Buonanno, \emph{{Conservative
  and radiative dynamics in classical relativistic scattering and bound
  systems}},  \href{https://arxiv.org/abs/2109.05994}{{\ttfamily 2109.05994}}.

\bibitem{HariDass:1980tq}
N.~D. Hari~Dass and V.~Soni, \emph{{Feynman Graph Derivation of Einstein
  Quadrupole Formula}},
  \href{https://doi.org/10.1088/0305-4470/15/2/019}{\emph{J. Phys. A}
  {\bfseries 15} (1982) 473}.

\bibitem{Thorne:1980ru}
K.~S. Thorne, \emph{{Multipole Expansions of Gravitational Radiation}},
  \href{https://doi.org/10.1103/RevModPhys.52.299}{\emph{Rev. Mod. Phys.}
  {\bfseries 52} (1980) 299}.

\bibitem{Blanchet:1985sp}
L.~Blanchet and T.~Damour, \emph{{Radiative gravitational fields in general
  relativity I. general structure of the field outside the source}},
  \href{https://doi.org/10.1098/rsta.1986.0125}{\emph{Phil. Trans. Roy. Soc.
  Lond.} {\bfseries A320} (1986) 379}.

\bibitem{Blanchet:1995fg}
L.~Blanchet, T.~Damour and B.~R. Iyer, \emph{{Gravitational waves from
  inspiralling compact binaries: Energy loss and wave form to second
  postNewtonian order}},
  \href{https://doi.org/10.1103/PhysRevD.51.5360}{\emph{Phys. Rev. D}
  {\bfseries 51} (1995) 5360}
  [\href{https://arxiv.org/abs/gr-qc/9501029}{{\ttfamily gr-qc/9501029}}].

\bibitem{Blanchet:1998in}
L.~Blanchet, \emph{{On the multipole expansion of the gravitational field}},
  \href{https://doi.org/10.1088/0264-9381/15/7/013}{\emph{Class. Quant. Grav.}
  {\bfseries 15} (1998) 1971}
  [\href{https://arxiv.org/abs/gr-qc/9801101}{{\ttfamily gr-qc/9801101}}].

\bibitem{Porto:2005ac}
R.~A. Porto, \emph{{Post-Newtonian corrections to the motion of spinning bodies
  in NRGR}}, \href{https://doi.org/10.1103/PhysRevD.73.104031}{\emph{Phys. Rev.
  D} {\bfseries 73} (2006) 104031}
  [\href{https://arxiv.org/abs/gr-qc/0511061}{{\ttfamily gr-qc/0511061}}].

\bibitem{Porto:2006bt}
R.~A. Porto and I.~Z. Rothstein, \emph{{The Hyperfine Einstein-Infeld-Hoffmann
  potential}}, \href{https://doi.org/10.1103/PhysRevLett.97.021101}{\emph{Phys.
  Rev. Lett.} {\bfseries 97} (2006) 021101}
  [\href{https://arxiv.org/abs/gr-qc/0604099}{{\ttfamily gr-qc/0604099}}].

\bibitem{Porto:2008tb}
R.~A. Porto and I.~Z. Rothstein, \emph{{Spin(1)Spin(2) Effects in the Motion of
  Inspiralling Compact Binaries at Third Order in the Post-Newtonian
  Expansion}}, \href{https://doi.org/10.1103/PhysRevD.78.044012}{\emph{Phys.
  Rev. D} {\bfseries 78} (2008) 044012}
  [\href{https://arxiv.org/abs/0802.0720}{{\ttfamily 0802.0720}}].

\bibitem{Levi:2015msa}
M.~Levi and J.~Steinhoff, \emph{{Spinning gravitating objects in the effective
  field theory in the post-Newtonian scheme}},
  \href{https://doi.org/10.1007/JHEP09(2015)219}{\emph{JHEP} {\bfseries 09}
  (2015) 219} [\href{https://arxiv.org/abs/1501.04956}{{\ttfamily
  1501.04956}}].

\bibitem{Levi:2014gsa}
M.~Levi and J.~Steinhoff, \emph{{Leading order finite size effects with spins
  for inspiralling compact binaries}},
  \href{https://doi.org/10.1007/JHEP06(2015)059}{\emph{JHEP} {\bfseries 06}
  (2015) 059} [\href{https://arxiv.org/abs/1410.2601}{{\ttfamily 1410.2601}}].

\bibitem{Levi:2018nxp}
M.~Levi, \emph{{Effective Field Theories of Post-Newtonian Gravity: A
  comprehensive review}},
  \href{https://doi.org/10.1088/1361-6633/ab12bc}{\emph{Rept. Prog. Phys.}
  {\bfseries 83} (2020) 075901}
  [\href{https://arxiv.org/abs/1807.01699}{{\ttfamily 1807.01699}}].

\bibitem{Levi:2016ofk}
M.~Levi and J.~Steinhoff, \emph{{Complete conservative dynamics for
  inspiralling compact binaries with spins at fourth post-Newtonian order}},
  \href{https://arxiv.org/abs/1607.04252}{{\ttfamily 1607.04252}}.

\bibitem{Tulczyjew:1959b}
W.~Tulczyjew, \emph{{Equations of motion of rotating bodies in general
  relativity theory}}, {\emph{Acta Phys.Polon.} {\bfseries 18} (1959) 37}.

\bibitem{Barker:1970zr}
B.~M. Barker and R.~F. O'Connell, \emph{{Derivation of the equations of motion
  of a gyroscope from the quantum theory of gravitation}},
  \href{https://doi.org/10.1103/PhysRevD.2.1428}{\emph{Phys. Rev.} {\bfseries
  D2} (1970) 1428}.

\bibitem{Barker:1975ae}
B.~M. Barker and R.~F. O'Connell, \emph{{Gravitational Two-Body Problem with
  Arbitrary Masses, Spins, and Quadrupole Moments}},
  \href{https://doi.org/10.1103/PhysRevD.12.329}{\emph{Phys. Rev.} {\bfseries
  D12} (1975) 329}.

\bibitem{D'Eath:1975vw}
P.~D. D'Eath, \emph{{Interaction of two black holes in the slow-motion limit}},
  \href{https://doi.org/10.1103/PhysRevD.12.2183}{\emph{Phys. Rev.} {\bfseries
  D12} (1975) 2183}.

\bibitem{Thorne:1984mz}
K.~S. Thorne and J.~B. Hartle, \emph{{Laws of motion and precession for black
  holes and other bodies}},
  \href{https://doi.org/10.1103/PhysRevD.31.1815}{\emph{Phys. Rev.} {\bfseries
  D31} (1984) 1815}.

\bibitem{Poisson:1997ha}
E.~Poisson, \emph{{Gravitational waves from inspiraling compact binaries: The
  Quadrupole moment term}},
  \href{https://doi.org/10.1103/PhysRevD.57.5287}{\emph{Phys. Rev.} {\bfseries
  D57} (1998) 5287} [\href{https://arxiv.org/abs/gr-qc/9709032}{{\ttfamily
  gr-qc/9709032}}].

\bibitem{Damour:2001tu}
T.~Damour, \emph{{Coalescence of two spinning black holes: an effective
  one-body approach}},
  \href{https://doi.org/10.1103/PhysRevD.64.124013}{\emph{Phys. Rev.}
  {\bfseries D64} (2001) 124013}
  [\href{https://arxiv.org/abs/gr-qc/0103018}{{\ttfamily gr-qc/0103018}}].

\bibitem{Hergt:2007ha}
S.~Hergt and G.~Schäfer, \emph{{Higher-order-in-spin interaction Hamiltonians
  for binary black holes from source terms of Kerr geometry in approximate ADM
  coordinates}}, \href{https://doi.org/10.1103/PhysRevD.77.104001}{\emph{Phys.
  Rev.} {\bfseries D77} (2008) 104001}
  [\href{https://arxiv.org/abs/0712.1515}{{\ttfamily 0712.1515}}].

\bibitem{Hergt:2008jn}
S.~Hergt and G.~Schäfer, \emph{{Higher-order-in-spin interaction Hamiltonians
  for binary black holes from Poincare invariance}},
  \href{https://doi.org/10.1103/PhysRevD.78.124004}{\emph{Phys. Rev.}
  {\bfseries D78} (2008) 124004}
  [\href{https://arxiv.org/abs/0809.2208}{{\ttfamily 0809.2208}}].

\bibitem{Vaidya:2014kza}
V.~Vaidya, \emph{{Gravitational spin Hamiltonians from the S matrix}},
  \href{https://doi.org/10.1103/PhysRevD.91.024017}{\emph{Phys. Rev.}
  {\bfseries D91} (2015) 024017}
  [\href{https://arxiv.org/abs/1410.5348}{{\ttfamily 1410.5348}}].

\bibitem{Marsat:2014xea}
S.~Marsat, \emph{{Cubic order spin effects in the dynamics and gravitational
  wave energy flux of compact object binaries}},
  \href{https://doi.org/10.1088/0264-9381/32/8/085008}{\emph{Class. Quant.
  Grav.} {\bfseries 32} (2015) 085008}
  [\href{https://arxiv.org/abs/1411.4118}{{\ttfamily 1411.4118}}].

\bibitem{Vines:2016qwa}
J.~Vines and J.~Steinhoff, \emph{{Spin-multipole effects in binary black holes
  and the test-body limit}},
  \href{https://arxiv.org/abs/1606.08832}{{\ttfamily 1606.08832}}.

\bibitem{Siemonsen:2017yux}
N.~Siemonsen, J.~Steinhoff and J.~Vines, \emph{{Gravitational waves from
  spinning binary black holes at the leading post-Newtonian orders at all
  orders in spin}},
  \href{https://doi.org/10.1103/PhysRevD.97.124046}{\emph{Phys. Rev. D}
  {\bfseries 97} (2018) 124046}
  [\href{https://arxiv.org/abs/1712.08603}{{\ttfamily 1712.08603}}].

\bibitem{Steinhoff:2014kwa}
J.~Steinhoff, \emph{{Spin and quadrupole contributions to the motion of
  astrophysical binaries}},
  \href{https://doi.org/10.1007/978-3-319-18335-0_19}{\emph{Fund. Theor. Phys.}
  {\bfseries 179} (2015) 615}
  [\href{https://arxiv.org/abs/1412.3251}{{\ttfamily 1412.3251}}].

\bibitem{Goldberger:2009qd}
W.~D. Goldberger and A.~Ross, \emph{{Gravitational radiative corrections from
  effective field theory}},
  \href{https://doi.org/10.1103/PhysRevD.81.124015}{\emph{Phys. Rev.}
  {\bfseries D81} (2010) 124015}
  [\href{https://arxiv.org/abs/0912.4254}{{\ttfamily 0912.4254}}].

\bibitem{Ross:2012fc}
A.~Ross, \emph{{Multipole expansion at the level of the action}},
  \href{https://doi.org/10.1103/PhysRevD.85.125033}{\emph{Phys. Rev.}
  {\bfseries D85} (2012) 125033}
  [\href{https://arxiv.org/abs/1202.4750}{{\ttfamily 1202.4750}}].

\bibitem{Harte:2016vwo}
A.~I. Harte and J.~Vines, \emph{{Generating exact solutions to Einstein’s
  equation using linearized approximations}},
  \href{https://doi.org/10.1103/PhysRevD.94.084009}{\emph{Phys. Rev.}
  {\bfseries D94} (2016) 084009}
  [\href{https://arxiv.org/abs/1608.04359}{{\ttfamily 1608.04359}}].

\bibitem{Vines:2017hyw}
J.~Vines, \emph{{Scattering of two spinning black holes in post-Minkowskian
  gravity, to all orders in spin, and effective-one-body mappings}},
  \href{https://arxiv.org/abs/1709.06016}{{\ttfamily 1709.06016}}.

\bibitem{Levi:2020kvb}
M.~Levi, A.~J. Mcleod and M.~Von~Hippel, \emph{{N$^3$LO gravitational
  spin-orbit coupling at order $G^4$}},
  \href{https://arxiv.org/abs/2003.02827}{{\ttfamily 2003.02827}}.

\bibitem{Levi:2020uwu}
M.~Levi, A.~J. Mcleod and M.~Von~Hippel, \emph{{NNNLO gravitational
  quadratic-in-spin interactions at the quartic order in G}},
  \href{https://arxiv.org/abs/2003.07890}{{\ttfamily 2003.07890}}.

\bibitem{Levi:2020lfn}
M.~Levi and F.~Teng, \emph{{NLO gravitational quartic-in-spin interaction}},
  \href{https://doi.org/10.1007/JHEP01(2021)066}{\emph{JHEP} {\bfseries 01}
  (2021) 066} [\href{https://arxiv.org/abs/2008.12280}{{\ttfamily
  2008.12280}}].

\bibitem{Siemonsen:2019dsu}
N.~Siemonsen and J.~Vines, \emph{{Test black holes, scattering amplitudes and
  perturbations of Kerr spacetime}},
  \href{https://doi.org/10.1103/PhysRevD.101.064066}{\emph{Phys. Rev. D}
  {\bfseries 101} (2020) 064066}
  [\href{https://arxiv.org/abs/1909.07361}{{\ttfamily 1909.07361}}].

\bibitem{Bernard:2015njp}
L.~Bernard, L.~Blanchet, A.~Bohé, G.~Faye and S.~Marsat, \emph{{Fokker action
  of nonspinning compact binaries at the fourth post-Newtonian approximation}},
  \href{https://doi.org/10.1103/PhysRevD.93.084037}{\emph{Phys. Rev.}
  {\bfseries D93} (2016) 084037}
  [\href{https://arxiv.org/abs/1512.02876}{{\ttfamily 1512.02876}}].

\bibitem{Buonanno:2012rv}
A.~Buonanno, G.~Faye and T.~Hinderer, \emph{{Spin effects on gravitational
  waves from inspiraling compact binaries at second post-Newtonian order}},
  \href{https://doi.org/10.1103/PhysRevD.87.044009}{\emph{Phys. Rev.}
  {\bfseries D87} (2013) 044009}
  [\href{https://arxiv.org/abs/1209.6349}{{\ttfamily 1209.6349}}].

\bibitem{Kidder:1992fr}
L.~E. Kidder, C.~M. Will and A.~G. Wiseman, \emph{{Spin effects in the inspiral
  of coalescing compact binaries}},
  \href{https://doi.org/10.1103/PhysRevD.47.R4183}{\emph{Phys. Rev.} {\bfseries
  D47} (1993) R4183} [\href{https://arxiv.org/abs/gr-qc/9211025}{{\ttfamily
  gr-qc/9211025}}].

\bibitem{Kidder:1995zr}
L.~E. Kidder, \emph{{Coalescing binary systems of compact objects to
  postNewtonian 5/2 order. 5. Spin effects}},
  \href{https://doi.org/10.1103/PhysRevD.52.821}{\emph{Phys. Rev.} {\bfseries
  D52} (1995) 821} [\href{https://arxiv.org/abs/gr-qc/9506022}{{\ttfamily
  gr-qc/9506022}}].

\bibitem{Misner1973}
C.~W. {Misner}, K.~S. {Thorne} and J.~A. {Wheeler}, \emph{{Gravitation}}. 1973.

\bibitem{Maggiore:1900zz}
M.~Maggiore, \emph{{Gravitational Waves. Vol. 1: Theory and Experiments}},
  Oxford Master Series in Physics. Oxford University Press, 2007.

\bibitem{Bern:2008qj}
Z.~Bern, J.~Carrasco and H.~Johansson, \emph{{New Relations for Gauge-Theory
  Amplitudes}}, \href{https://doi.org/10.1103/PhysRevD.78.085011}{\emph{Phys.
  Rev. D} {\bfseries 78} (2008) 085011}
  [\href{https://arxiv.org/abs/0805.3993}{{\ttfamily 0805.3993}}].

\bibitem{Bjerrum-Bohr:2013bxa}
N.~E.~J. Bjerrum-Bohr, J.~F. Donoghue and P.~Vanhove, \emph{{On-shell
  Techniques and Universal Results in Quantum Gravity}},
  \href{https://doi.org/10.1007/JHEP02(2014)111}{\emph{JHEP} {\bfseries 02}
  (2014) 111} [\href{https://arxiv.org/abs/1309.0804}{{\ttfamily 1309.0804}}].

\bibitem{Li:2018qap}
J.~Li and S.~G. Prabhu, \emph{{Gravitational radiation from the classical
  spinning double copy}},
  \href{https://doi.org/10.1103/PhysRevD.97.105019}{\emph{Phys. Rev. D}
  {\bfseries 97} (2018) 105019}
  [\href{https://arxiv.org/abs/1803.02405}{{\ttfamily 1803.02405}}].

\bibitem{Goldberger:2016iau}
W.~D. Goldberger and A.~K. Ridgway, \emph{{Radiation and the classical double
  copy for color charges}},
  \href{https://doi.org/10.1103/PhysRevD.95.125010}{\emph{Phys. Rev. D}
  {\bfseries 95} (2017) 125010}
  [\href{https://arxiv.org/abs/1611.03493}{{\ttfamily 1611.03493}}].

\bibitem{Luna:2017dtq}
A.~Luna, I.~Nicholson, D.~O'Connell and C.~D. White, \emph{{Inelastic Black
  Hole Scattering from Charged Scalar Amplitudes}},
  \href{https://doi.org/10.1007/JHEP03(2018)044}{\emph{JHEP} {\bfseries 03}
  (2018) 044} [\href{https://arxiv.org/abs/1711.03901}{{\ttfamily
  1711.03901}}].

\bibitem{Peters:1970mx}
P.~C. Peters, \emph{{Relativistic gravitational bremsstrahlung}},
  \href{https://doi.org/10.1103/PhysRevD.1.1559}{\emph{Phys. Rev. D} {\bfseries
  1} (1970) 1559}.

\bibitem{1985Konradin}
K.~Westpfahl, \emph{High-speed scattering of charged and uncharged particles in
  general relativity}, .

\bibitem{1977KT}
S.~J. Kovacs and K.~S. Thorne, \emph{The generation of gravitational waves.
  {III} - derivation of bremsstrahlung formulae}, .

\bibitem{1978KT}
J.~K. S.~J. and K.~S. Thorne, \emph{The generation of gravitational waves. {IV}
  - bremsstrahlung}, .

\end{thebibliography}\endgroup
\end{document}